\documentclass[11pt]{article}
\input{psfig.sty}
\newcommand{\dshv}{
\begin{picture}(52,10)
\put(15,3){\line(-5,2){10}}
\put(15,3){\line(-5,-2){10}}
\multiput(15,3)(22,0){2}{\circle*{2}}
\multiput(15,3)(6,0){4}{\line(1,0){4}}
\put(37,3){\line(5,2){10}}
\put(37,3){\line(5,-2){10}}
\end{picture}  }
\newcommand{\selfe}{
\begin{picture}(48,10)
\thicklines
\put(24,3.5){\circle{11}}
\multiput(23.8,-1.6)(0.5,0.5){12}{\circle*{1}}
\multiput(24.2,8.6)(-0.5,-0.5){12}{\circle*{1}}
\multiput(21.2,-1)(0.5,0.5){15}{\circle*{1}}
\multiput(26.8,8)(-0.5,-0.5){15}{\circle*{1}}
\put(18.5,3.5){\line(-1,0){15}}
\put(29.5,3.5){\line(1,0){15}}
\end{picture}  }
\newcommand{\skd}{
\begin{picture}(45,10)
\multiput(9,3.5)(27,0){2}{\circle{11}}
\multiput(14.5,3.5)(16,0){2}{\circle*{3}}
\multiput(14.5,3.5)(6,0){3}{\line(1,0){4}}
\end{picture}  }
\newcommand{\dg}{
\begin{picture}(55,14)
\put(27.5,3.5){\circle{16}}
\multiput(19.5,3.5)(16,0){2}{\circle*{3}}
\multiput(19.5,3.5)(-6,0){3}{\line(-1,0){4}}
\multiput(35.5,3.5)(6,0){3}{\line(1,0){4}}
\put(3.5,7.5){\footnotesize {\bf q}}
\put(43,7.5){\footnotesize -{\bf q}}
\end{picture}  }
\newcommand{\dgn}{
\begin{picture}(55,14)
\put(27.5,3.5){\circle{16}}
\multiput(19.5,3.5)(16,0){2}{\circle*{3}}
\multiput(19.5,3.5)(-6,0){3}{\line(-1,0){4}}
\multiput(35.5,3.5)(6,0){3}{\line(1,0){4}}
\put(5,6){\footnotesize {\bf 0}}
\put(44.5,6){\footnotesize {\bf 0}}
\end{picture}  }
\newcommand{\dgbn}{
\begin{picture}(69,14)
\put(26.5,3.5){\oval(14,14)[l]}
\put(42.5,3.5){\oval(14,14)[r]}
\multiput(19.5,3.5)(30,0){2}{\circle*{3}}
\multiput(19.5,3.5)(-6,0){3}{\line(-1,0){4}}
\multiput(49.5,3.5)(6,0){3}{\line(1,0){4}}
\put(5,6){\footnotesize {\bf 0}}
\put(58.5,6){\footnotesize {\bf 0}}
\put(26.5,-3.5){\line(1,0){16}}
\put(26.5,10.5){\line(1,0){16}}
\put(27,10.5){\line(0,-1){14}}
\put(43,10.5){\line(0,-1){14}}
\bezier{50}(27,6)(29.25,8.25)(31.5,10.5)
\bezier{100}(27,-1)(32.75,4.75)(38.5,10.5)
\bezier{100}(31.5,-3.5)(37.25,2.25)(43,8)
\bezier{50}(38.5,-3.5)(40.75,-1.25)(43,1)
\bezier{50}(31.5,-3.5)(29.25,-1.25)(27,1)
\bezier{100}(38.5,-3.5)(32.75,2.25)(27,8)
\bezier{100}(31.5,10.5)(37.25,4.75)(43,-1)
\bezier{50}(38.5,10.5)(40.75,8.25)(43,6)
\end{picture}  }
\newcommand{\sixz}{
\begin{picture}(57,23)
{\thicklines
\put(20,2){\line(1,0){16}}
\put(20,16){\line(1,0){16}}
\put(19.5,16){\line(0,-1){14}}
\put(36.5,16){\line(0,-1){14}}
\put(20,2){\line(-3,-1){10}}
\put(20,16){\line(-3,1){10}}
\put(36,2){\line(3,-1){10}}
\put(36,16){\line(3,1){10}}
\put(20,9){\line(-1,0){10}}
\put(36,9){\line(1,0){10}}
\put(3,-2){\footnotesize {\bf 0}}
\put(3,14.5){\footnotesize {\bf 0}}
\put(48,-2){\footnotesize {\bf 0}}
\put(48,14.5){\footnotesize {\bf 0}}
\put(3,6.25){\footnotesize {\bf 0}}
\put(48,6.25){\footnotesize {\bf 0}}
}
\bezier{50}(20,11.5)(22.25,13.75)(24.5,16)
\bezier{100}(20,4.5)(25.75,10.25)(31.5,16)
\bezier{100}(24.5,2)(30.25,7.75)(36,13.5)
\bezier{50}(31.5,2)(33.75,4.25)(36,6.5)
\end{picture}  }
\newcommand{\fourz}{
\begin{picture}(57,18)
{\thicklines
\put(10,-1){\line(1,0){36}}
\put(10,13){\line(1,0){36}}
\put(19.5,13){\line(0,-1){14}}
\put(36.5,13){\line(0,-1){14}}
\put(3,-1){\footnotesize {\bf 0}}
\put(3,10){\footnotesize {\bf 0}}
\put(48,-1){\footnotesize {\bf 0}}
\put(48,10){\footnotesize {\bf 0}}
}
\bezier{50}(20,8.5)(22.25,10.75)(24.5,13)
\bezier{100}(20,1.5)(25.75,7.25)(31.5,13)
\bezier{100}(24.5,-1)(30.25,4.75)(36,10.5)
\bezier{50}(31.5,-1)(33.75,1.25)(36,3.5)
\end{picture}  }
\newcommand{\fourzz}{
\begin{picture}(57,18)
{\thicklines
\put(10,-1){\line(1,0){36}}
\put(10,13){\line(1,0){36}}
\put(19.5,13){\line(0,-1){14}}
\put(36.5,13){\line(0,-1){14}}
\put(3,-2){\footnotesize {\bf 0}}
\put(3,9){\footnotesize {\bf 0}}
\put(48,-2){\footnotesize {\bf 0}}
\put(48,9){\footnotesize {\bf 0}}
}
\bezier{50}(20,8.5)(22.25,10.75)(24.5,13)
\bezier{100}(20,1.5)(25.75,7.25)(31.5,13)
\bezier{100}(24.5,-1)(30.25,4.75)(36,10.5)
\bezier{50}(31.5,-1)(33.75,1.25)(36,3.5)
\bezier{50}(24.5,-1)(22.25,1.25)(20,3.5)
\bezier{100}(31.5,-1)(25.75,4.75)(20,10.5)
\bezier{100}(24.5,13)(30.25,7.25)(36,1.5)
\bezier{50}(31.5,13)(33.75,10.75)(36,8.5)
\end{picture}  }
\newcommand{\four}{
\begin{picture}(45,16)
{\thicklines
\put(4,-1){\line(1,0){36}}
\put(4,13){\line(1,0){36}}
\put(13.5,13){\line(0,-1){14}}
\put(30.5,13){\line(0,-1){14}}
}
\bezier{50}(14,8.5)(16.25,10.75)(18.5,13)
\bezier{100}(14,1.5)(19.75,7.25)(25.5,13)
\bezier{100}(18.5,-1)(24.25,4.75)(30,10.5)
\bezier{50}(25.5,-1)(27.75,1.25)(30,3.5)
\bezier{50}(18.5,-1)(16.25,1.25)(14,3.5)
\bezier{100}(25.5,-1)(19.75,4.75)(14,10.5)
\bezier{100}(18.5,13)(24.25,7.25)(30,1.5)
\bezier{50}(25.5,13)(27.75,10.75)(30,8.5)
\end{picture}  }
\newcommand{\fourt}{
\begin{picture}(90,16)
\put(4,-1){\line(1,0){26}}
\put(4,13){\line(1,0){26}}
\put(14,13){\line(0,-1){14}}
\put(30,13){\line(0,-1){14}}
\bezier{50}(14,8.5)(16.25,10.75)(18.5,13)
\bezier{100}(14,1.5)(19.75,7.25)(25.5,13)
\bezier{100}(18.5,-1)(24.25,4.75)(30,10.5)
\bezier{50}(25.5,-1)(27.75,1.25)(30,3.5)
\bezier{50}(18.5,-1)(16.25,1.25)(14,3.5)
\bezier{100}(25.5,-1)(19.75,4.75)(14,10.5)
\bezier{100}(18.5,13)(24.25,7.25)(30,1.5)
\bezier{50}(25.5,13)(27.75,10.75)(30,8.5)
\put(30.5,6){\oval(14,14)[r]}
\multiput(39.5,6)(8,0){2}{\oval(4,4)[t]}
\multiput(43.5,6)(8,0){2}{\oval(4,4)[b]}
\multiput(37.5,6)(16,0){2}{\circle*{2}}
\put(60.5,6){\oval(14,14)[l]}
\put(60,-1){\line(1,0){26}}
\put(60,13){\line(1,0){26}}
\put(60,13){\line(0,-1){14}}
\put(76,13){\line(0,-1){14}}
\bezier{50}(60,8.5)(62.25,10.75)(64.5,13)
\bezier{100}(60,1.5)(65.75,7.25)(71.5,13)
\bezier{100}(64.5,-1)(70.25,4.75)(76,10.5)
\bezier{50}(71.5,-1)(73.75,1.25)(76,3.5)
\bezier{50}(64.5,-1)(62.25,1.25)(60,3.5)
\bezier{100}(71.5,-1)(65.75,4.75)(60,10.5)
\bezier{100}(64.5,13)(70.25,7.25)(76,1.5)
\bezier{50}(71.5,13)(73.75,10.75)(76,8.5)
\end{picture}  }
\newcommand{\fourtz}{
\begin{picture}(103,18)
\put(10,-1){\line(1,0){26}}
\put(10,13){\line(1,0){26}}
\put(20,13){\line(0,-1){14}}
\put(36,13){\line(0,-1){14}}
\bezier{50}(20,8.5)(22.25,10.75)(24.5,13)
\bezier{100}(20,1.5)(25.75,7.25)(31.5,13)
\bezier{100}(24.5,-1)(30.25,4.75)(36,10.5)
\bezier{50}(31.5,-1)(33.75,1.25)(36,3.5)
\bezier{50}(24.5,-1)(22.25,1.25)(20,3.5)
\bezier{100}(31.5,-1)(25.75,4.75)(20,10.5)
\bezier{100}(24.5,13)(30.25,7.25)(36,1.5)
\bezier{50}(31.5,13)(33.75,10.75)(36,8.5)
\put(36.5,6){\oval(14,14)[r]}
\multiput(45.5,6)(8,0){2}{\oval(4,4)[t]}
\multiput(49.5,6)(8,0){2}{\oval(4,4)[b]}
\multiput(43.5,6)(16,0){2}{\circle*{2}}
\put(66.5,6){\oval(14,14)[l]}
\put(66,-1){\line(1,0){26}}
\put(66,13){\line(1,0){26}}
\put(66,13){\line(0,-1){14}}
\put(82,13){\line(0,-1){14}}
\bezier{50}(66,8.5)(68.25,10.75)(70.5,13)
\bezier{100}(66,1.5)(71.75,7.25)(77.5,13)
\bezier{100}(70.5,-1)(76.25,4.75)(82,10.5)
\bezier{50}(77.5,-1)(79.75,1.25)(82,3.5)
\bezier{50}(70.5,-1)(68.25,1.25)(66,3.5)
\bezier{100}(77.5,-1)(71.75,4.75)(66,10.5)
\bezier{100}(70.5,13)(76.25,7.25)(82,1.5)
\bezier{50}(77.5,13)(79.75,10.75)(82,8.5)
\put(3,-2){\footnotesize {\bf 0}}
\put(3,9){\footnotesize {\bf 0}}
\put(94,-2){\footnotesize {\bf 0}}
\put(94,9){\footnotesize {\bf 0}}
\end{picture}  }
\newcommand{\fourkz}{
\begin{picture}(57,18)
{\thicklines
\put(10,-1){\line(1,0){36}}
\put(10,13){\line(1,0){36}}
\put(19.5,13){\line(0,-1){14}}
\put(36.5,13){\line(0,-1){14}}
\put(0,-1){\footnotesize -{\bf k}}
\put(3,10){\footnotesize {\bf k}}
\put(48,-1){\footnotesize {\bf 0}}
\put(48,10){\footnotesize {\bf 0}}
}
\bezier{50}(20,8.5)(22.25,10.75)(24.5,13)
\bezier{100}(20,1.5)(25.75,7.25)(31.5,13)
\bezier{100}(24.5,-1)(30.25,4.75)(36,10.5)
\bezier{50}(31.5,-1)(33.75,1.25)(36,3.5)
\bezier{50}(24.5,-1)(22.25,1.25)(20,3.5)
\bezier{100}(31.5,-1)(25.75,4.75)(20,10.5)
\bezier{100}(24.5,13)(30.25,7.25)(36,1.5)
\bezier{50}(31.5,13)(33.75,10.75)(36,8.5)
\end{picture}  }
\newcommand{\fourkq}{
\begin{picture}(60,18)
{\thicklines
\put(10,-1){\line(1,0){36}}
\put(10,13){\line(1,0){36}}
\put(19.5,13){\line(0,-1){14}}
\put(36.5,13){\line(0,-1){14}}
\put(0,-1){\footnotesize -{\bf k}}
\put(3,10){\footnotesize {\bf k}}
\put(48,-1){\footnotesize -{\bf q}}
\put(48,10){\footnotesize {\bf q}}
}
\bezier{50}(20,8.5)(22.25,10.75)(24.5,13)
\bezier{100}(20,1.5)(25.75,7.25)(31.5,13)
\bezier{100}(24.5,-1)(30.25,4.75)(36,10.5)
\bezier{50}(31.5,-1)(33.75,1.25)(36,3.5)
\bezier{50}(24.5,-1)(22.25,1.25)(20,3.5)
\bezier{100}(31.5,-1)(25.75,4.75)(20,10.5)
\bezier{100}(24.5,13)(30.25,7.25)(36,1.5)
\bezier{50}(31.5,13)(33.75,10.75)(36,8.5)
\end{picture}  }
\newcommand{\ddg}{
\begin{picture}(63,17)
\put(31.5,3.5){\circle{24}}
\multiput(19.5,3.5)(24,0){2}{\circle*{3}}
\multiput(19.5,3.5)(-6,0){3}{\line(-1,0){4}}
\multiput(43.5,3.5)(6,0){3}{\line(1,0){4}}
\put(3.5,7.5){\footnotesize {\bf q}}
\put(50.5,7.5){\footnotesize -{\bf q}}
\multiput(31.5,15.5)(0,-24){2}{\circle*{3}}
\multiput(31.7,12.3)(0,-7.2){3}{\oval(4,4)[tl]}
\multiput(31.7,12.7)(0,-7.2){3}{\oval(4,4)[bl]}
\multiput(31.3,8.7)(0,-7.2){3}{\oval(4,4)[tr]}
\multiput(31.3,9.1)(0,-7.2){3}{\oval(4,4)[br]}
\end{picture}  }
\newcommand{\dddg}{
\begin{picture}(79,17)
\put(39.5,3.5){\oval(40,24)}
\multiput(19.5,3.5)(40,0){2}{\circle*{3}}
\multiput(19.5,3.5)(-6,0){3}{\line(-1,0){4}}
\multiput(59.5,3.5)(6,0){3}{\line(1,0){4}}
\put(3.5,7.5){\footnotesize {\bf q}}
\put(66.5,7.5){\footnotesize -{\bf q}}
\multiput(31.5,15.5)(0,-24){2}{\circle*{3}}
\multiput(31.7,12.3)(0,-7.2){3}{\oval(4,4)[tl]}
\multiput(31.7,12.7)(0,-7.2){3}{\oval(4,4)[bl]}
\multiput(31.3,8.7)(0,-7.2){3}{\oval(4,4)[tr]}
\multiput(31.3,9.1)(0,-7.2){3}{\oval(4,4)[br]}
\multiput(47.5,15.5)(0,-24){2}{\circle*{3}}
\multiput(47.7,12.3)(0,-7.2){3}{\oval(4,4)[tl]}
\multiput(47.7,12.7)(0,-7.2){3}{\oval(4,4)[bl]}
\multiput(47.3,8.7)(0,-7.2){3}{\oval(4,4)[tr]}
\multiput(47.3,9.1)(0,-7.2){3}{\oval(4,4)[br]}
\end{picture}  }
\newcommand{\dddgg}{
\begin{picture}(87,17)
\put(43.5,3.5){\oval(48,24)}
\multiput(19.5,3.5)(48,0){2}{\circle*{3}}
\multiput(19.5,3.5)(-6,0){3}{\line(-1,0){4}}
\multiput(67.5,3.5)(6,0){3}{\line(1,0){4}}
\put(3.5,7.5){\footnotesize {\bf q}}
\put(74.5,7.5){\footnotesize -{\bf q}}
\multiput(31.5,15.5)(0,-24){2}{\circle*{3}}
\put(31.5,15.5){\line(0,-1){2}}
\multiput(35.5,13.9)(8,-7.2){3}{\oval(8,8)[bl]}
\multiput(35.5,5.9)(8,-7.2){3}{\oval(8,8)[tr]}
\multiput(55.5,15.5)(0,-24){2}{\circle*{3}}
\multiput(51.5,15.5)(-8,-7.2){3}{\oval(8,8)[br]}
\multiput(51.5,7.5)(-8,-7.2){3}{\oval(8,8)[tl]}
\put(31.5,-8.5){\line(0,1){2}}
\end{picture}  }
\newcommand{\dddggg}{
\begin{picture}(67,19)
\put(33.5,3.5){\circle{28}}
\multiput(19.5,3.5)(28,0){2}{\circle*{3}}
\multiput(19.5,3.5)(-6,0){3}{\line(-1,0){4}}
\multiput(47.5,3.5)(6,0){3}{\line(1,0){4}}
\put(3.5,7.5){\footnotesize {\bf q}}
\put(54.5,7.5){\footnotesize -{\bf q}}
\multiput(33.5,17.5)(0,-28){2}{\circle*{3}}
\multiput(33.7,14.1)(0,-7.2){4}{\oval(4,4)[tl]}
\multiput(33.7,14.5)(0,-7.2){4}{\oval(4,4)[bl]}
\multiput(33.3,10.5)(0,-7.2){3}{\oval(4,4)[tr]}
\multiput(33.3,10.9)(0,-7.2){3}{\oval(4,4)[br]}
\multiput(27.75,-2.5)(8,0){3}{\oval(4,4)[t]}
\multiput(23.75,-2.1)(8,0){3}{\oval(4,4)[b]}
\multiput(20.85,-2.5)(25.3,0){2}{\circle*{3}}
\end{picture}  }
\newcommand{\dddgggg}{
\begin{picture}(93.14,15)
\multiput(29.5,3.5)(34.14,0){2}{\circle{20}}
\multiput(19.5,3.5)(54.14,0){2}{\circle*{3}}
\multiput(19.5,3.5)(-6,0){3}{\line(-1,0){4}}
\multiput(73.64,3.5)(6,0){3}{\line(1,0){4}}
\put(3.5,7.5){\footnotesize {\bf q}}
\put(80.64,7.5){\footnotesize -{\bf q}}
\multiput(36.57,10.57)(20,0){2}{\circle*{3}}
\multiput(36.57,-3.57)(20,0){2}{\circle*{3}}
\multiput(38.57,10.57)(8,0){3}{\oval(4,4)[t]}
\multiput(42.57,10.57)(8,0){2}{\oval(4,4)[b]}
\multiput(38.57,-3.57)(8,0){3}{\oval(4,4)[b]}
\multiput(42.57,-3.57)(8,0){2}{\oval(4,4)[t]}
\end{picture}  }
\newcommand{\two}{
\begin{picture}(77.8,10)
\multiput(25,3.5)(27.8,0){2}{\circle{11}}
\multiput(24.8,-1.6)(0.5,0.5){12}{\circle*{1}}
\multiput(25.2,8.6)(-0.5,-0.5){12}{\circle*{1}}
\multiput(22.2,-1)(0.5,0.5){15}{\circle*{1}}
\multiput(27.8,8)(-0.5,-0.5){15}{\circle*{1}}
\multiput(52.6,-1.6)(0.5,0.5){12}{\circle*{1}}
\multiput(53,8.6)(-0.5,-0.5){12}{\circle*{1}}
\multiput(50,-1)(0.5,0.5){15}{\circle*{1}}
\multiput(55.6,8)(-0.5,-0.5){15}{\circle*{1}}
\multiput(19.5,3.5)(-6,0){3}{\line(-1,0){4}}
\multiput(58.3,3.5)(6,0){3}{\line(1,0){4}}
\multiput(32.9,3.5)(8,0){2}{\oval(4,4)[t]}
\multiput(36.9,3.5)(8,0){2}{\oval(4,4)[b]}
\end{picture}  }
\newcommand{\wblock}{
\begin{picture}(50,10)
\put(25,3.5){\circle{11}}
\multiput(24.8,-1.6)(0.5,0.5){12}{\circle*{1}}
\multiput(25.2,8.6)(-0.5,-0.5){12}{\circle*{1}}
\multiput(22.2,-1)(0.5,0.5){15}{\circle*{1}}
\multiput(27.8,8)(-0.5,-0.5){15}{\circle*{1}}
\multiput(17.1,3.5)(-8,0){2}{\oval(4,4)[t]}
\multiput(13.1,3.5)(-8,0){2}{\oval(4,4)[b]}
\multiput(32.9,3.5)(8,0){2}{\oval(4,4)[t]}
\multiput(36.9,3.5)(8,0){2}{\oval(4,4)[b]}
\end{picture}  }
\newcommand{\wl}{
\begin{picture}(50,12)
\multiput(7,0)(8,0){5}{\oval(4,4)[t]}
\multiput(11,0)(8,0){5}{\oval(4,4)[b]}
\put(5,3.5){\footnotesize {\bf k}}
\put(37,3.5){\footnotesize -{\bf k}}
\end{picture}  }
\newcommand{\wlpart}{
\begin{picture}(36,10)
\put(21,3){\circle*{2}}
\put(21,3){\line(5,2){10}}
\put(21,3){\line(5,-2){10}}
\multiput(7,3)(8,0){2}{\oval(4,4)[t]}
\multiput(11,3)(8,0){2}{\oval(4,4)[b]}
\end{picture} }

\newcommand{\blocks}{
\begin{picture}(80,18)
\multiput(26.5,7)(27,0){2}{\circle{11}}
\multiput(5,7)(6,0){3}{\line(1,0){4}}
\multiput(59,7)(6,0){3}{\line(1,0){4}}
\multiput(26.3,1.9)(0.5,0.5){12}{\circle*{1}}
\multiput(26.7,12.1)(-0.5,-0.5){12}{\circle*{1}}
\multiput(23.7,2.5)(0.5,0.5){15}{\circle*{1}}
\multiput(29.3,11.5)(-0.5,-0.5){15}{\circle*{1}}
\multiput(53.3,1.9)(0.5,0.5){12}{\circle*{1}}
\multiput(53.7,12.1)(-0.5,-0.5){12}{\circle*{1}}
\multiput(50.7,2.5)(0.5,0.5){15}{\circle*{1}}
\multiput(56.3,11.5)(-0.5,-0.5){15}{\circle*{1}}
\bezier{100}(32,9)(40,11.5)(48,9)
\bezier{100}(32,5)(40,2.5)(48,5)
\bezier{100}(30,11.5)(40,18.5)(50,11.5)
\bezier{100}(30,2.5)(40,-4.5)(50,2.5)
\end{picture} }
\newcommand{\blockn}{
\begin{picture}(80,18)
\multiput(26.5,7)(27,0){2}{\circle{11}}
\multiput(5,7)(6,0){3}{\line(1,0){4}}
\multiput(59,7)(6,0){3}{\line(1,0){4}}
\multiput(26.3,1.9)(0.5,0.5){12}{\circle*{1}}
\multiput(26.7,12.1)(-0.5,-0.5){12}{\circle*{1}}
\multiput(23.7,2.5)(0.5,0.5){15}{\circle*{1}}
\multiput(29.3,11.5)(-0.5,-0.5){15}{\circle*{1}}
\multiput(53.3,1.9)(0.5,0.5){12}{\circle*{1}}
\multiput(53.7,12.1)(-0.5,-0.5){12}{\circle*{1}}
\multiput(50.7,2.5)(0.5,0.5){15}{\circle*{1}}
\multiput(56.3,11.5)(-0.5,-0.5){15}{\circle*{1}}
\bezier{100}(32,9)(40,11.5)(48,9)
\bezier{100}(32,5)(40,2.5)(48,5)
\bezier{100}(30,11.5)(40,18.5)(50,11.5)
\bezier{100}(30,2.5)(40,-4.5)(50,2.5)
\put(8,9){\footnotesize {\bf 0}}
\put(67,9){\footnotesize {\bf 0}}
\end{picture} }
\newcommand{\ssb}{
\begin{picture}(50,10)
\put(25,3.5){\circle{11}}
\multiput(24.8,-1.6)(0.5,0.5){12}{\circle*{1}}
\multiput(25.2,8.6)(-0.5,-0.5){12}{\circle*{1}}
\multiput(22.2,-1)(0.5,0.5){15}{\circle*{1}}
\multiput(27.8,8)(-0.5,-0.5){15}{\circle*{1}}
\multiput(19.5,3.5)(-6,0){3}{\line(-1,0){4}}
\multiput(30.5,3.5)(6,0){3}{\line(1,0){4}}
\end{picture} }
\newcommand{\selfek}{
\begin{picture}(54,14)
\thicklines
\put(26,3.5){\circle{11}}
\multiput(25.8,-1.6)(0.5,0.5){12}{\circle*{1}}
\multiput(26.2,8.6)(-0.5,-0.5){12}{\circle*{1}}
\multiput(23.2,-1)(0.5,0.5){15}{\circle*{1}}
\multiput(28.8,8)(-0.5,-0.5){15}{\circle*{1}}
\put(20.5,3.5){\line(-1,0){17}}
\put(31.5,3.5){\line(1,0){19}}
\put(3.5,5.5){\footnotesize {\bf k}}
\put(42,5.5){\footnotesize -{\bf k}}
\multiput(12,6.5)(0.25,-0.5){13}{\circle*{1}}
\multiput(12,0.5)(0.25,0.5){13}{\circle*{1}}
\multiput(40,6.5)(-0.25,-0.5){13}{\circle*{1}}
\multiput(40,0.5)(-0.25,0.5){13}{\circle*{1}}
\end{picture} }
\newcommand{\skedd}{
\begin{picture}(100,18)
\multiput(30.5,-1)(36,0){2}{\circle*{3}}
{\thicklines
\put(30.5,-1){\line(-1,0){27}}
\put(66.5,-1){\line(1,0){30}}}
\put(3.5,1){i,{\bf k}}
\put(80,1){i,-{\bf k}}
\multiput(22,2)(0.25,-0.5){13}{\circle*{1}}
\multiput(22,-4)(0.25,0.5){13}{\circle*{1}}
\multiput(75,2)(-0.25,-0.5){13}{\circle*{1}}
\multiput(75,-4)(-0.25,0.5){13}{\circle*{1}}
\put(48.5,-1){\oval(36,36)[t]}
\multiput(32.5,-1)(8,0){5}{\oval(4,4)[b]}
\multiput(36.5,-1)(8,0){4}{\oval(4,4)[t]}
\end{picture} }
\newcommand{\skeddd}{
\begin{picture}(100,18)
\multiput(30.5,-1)(36,0){2}{\circle*{3}}
{\thicklines
\put(30.5,-1){\line(-1,0){27}}
\put(66.5,-1){\line(1,0){30}}}
\put(3.5,1){i,{\bf k}}
\put(80,1){i,-{\bf k}}
\multiput(22,2)(0.25,-0.5){13}{\circle*{1}}
\multiput(22,-4)(0.25,0.5){13}{\circle*{1}}
\multiput(75,2)(-0.25,-0.5){13}{\circle*{1}}
\multiput(75,-4)(-0.25,0.5){13}{\circle*{1}}
\put(48.5,-1){\oval(36,36)[t]}
\multiput(43.7,13)(7.7,-4){4}{\oval(8,4)[bl]}
\multiput(43.4,9)(7.7,-4){3}{\oval(8,4)[tr]}
\multiput(53.3,13)(-7.7,-4){4}{\oval(8,4)[br]}
\multiput(53.6,9)(-7.7,-4){3}{\oval(8,4)[tl]}
\put(39.2,14.5){\circle*{3}}
\put(57.8,14.5){\circle*{3}}
\end{picture} }
\begin{document}

\title{
\textbf{Longitudinal and transverse Greens
functions in $\varphi^4$ model below and near the critical point}
}

\author{J. Kaupu\v{z}s
\thanks{E--mail: \texttt{kaupuzs@latnet.lv}} \\
Institute of Mathematics and Computer Science, University of Latvia\\
29 Rainja Boulevard, LV--1459 Riga, Latvia}

\date{\today}

\maketitle

\begin{abstract}
We have extended our method of grouping of Feynman diagrams (GFD theory)
to study the transverse and longitudinal Greens functions
$G_{\perp}({\bf k})$ and $G_{\parallel}({\bf k})$
in $\varphi^4$ model below the critical point ($T<T_c$)
in presence of an infinitesimal external field.
Our method allows a qualitative analysis not cutting the perturbation 
series. We have shown that the critical behavior
of the Greens functions is consistent with a general scaling hypothesis,
where the same critical exponents, found within the GFD theory,
are valid both at $T<T_c$ and $T>T_c$. The long--wave limit $k \to 0$
has been studied at $T<T_c$, showing that 
$G_{\perp}({\bf k}) \simeq a \, k^{-\lambda_{\perp}}$ and
$G_{\parallel}({\bf k}) \simeq b \, k^{-\lambda_{\parallel}}$
with exponents $d/2<\lambda_{\perp}<2$ and $\lambda_{\parallel}
=2 \lambda_{\perp} -d$ is the physical solution of our equations
at the spatial dimensionality $2<d<4$, which coincides with the asymptotic
solution at $T \to T_c$ as well as with a non--perturbative
renormalization group (RG) analysis provided in our paper.
It is confirmed also by Monte Carlo simulation.
The exponents, as well as the ratio $b M^2/a^2$ (where $M$ is magnetization)
are universal.
The results of the perturbative RG method
are reproduced by formally setting $\lambda_{\perp}=2$.
Nevertheless, we disprove the conventional statement that
$\lambda_{\perp}=2$ is the exact result.
\end{abstract}

\section{Introduction}

Phase transitions and critical phenomena is a widely investigated
field in physics and natural sciences~\cite{Sornette,BDT,Ma,Justin}.
The current paper is devoted to further development of our original
diagrammatic method introduced in~\cite{K1}, to study the $\varphi^4$
phase transition model below the critical point. Our approach is
based on a suitable grouping of Feynman diagrams, therefore we
shall call it the GFD theory. In distinction to the conventional
perturbative renormalization group (RG) method~\cite{Ma,Justin}, it
allows a qualitative analysis near and at the critical point,
not cutting the perturbation series. In such a way, we have found the
set of possible values for exact critical exponents~\cite{K1} in two and
three dimensions in agreement with the known exact results
for the two--dimensional Ising model~\cite{Onsager,Baxter}.
A good agreement with some Monte Carlo (MC) data~\cite{IS,SM} and
experiments~\cite{GA} has been found in~\cite{K1}, as well.
The disagreement with the conventionally accepted RG values of the
critical exponents has been widely discussed
in~\cite{K2,K3}, providing arguments that very sensitive numerical tests
confirm our theoretical predictions. 

The $\varphi^4$ model exhibits a nontrivial
behavior in close vicinity, as well as below the critical temperature
$T_c$, if the order parameter is an $n$--component vector
with $n>1$. The related long--wave divergence of the longitudinal
and transverse correlation functions (in Fourier representation)
at $T<T_c$ has been studied in~\cite{PaPo,SchwMi}
based on the hydrodynamical (Gaussian) approximation.
Essentially the same problem has been studied before in~\cite{Wagner}
in terms of the Gaussian spin--wave theory~\cite{Dyson}.
Later perturbative renormalization group (RG)
studies~\cite{FBJ,BW,WZ,Nel,BZ,Sch,Law1,Law2,Tu}
support the Gaussian approximation.
The RG method is claimed to be asymptotically exact.
However, we disprove this statement by finding out the errors
as well as unjustified assertions and assumptions in the papers where these 
claims have been established (Sec.~\ref{sec:cran}). 
Our analysis predicts
a non--Gaussian behavior, and we show by general physical
arguments that it must be the true behavior to coincide
with the know rigorous results for the classical $XY$ model. 
This prediction is strongly supported by a Monte Carlo
test (Sec.~\ref{sec:test}).

\section{Primary equations} \label{sec:pri}

  We consider a $\varphi^4$ model with the Hamiltonian
\begin{equation} \label{eq:H}
H/T= \int \left( r_0 \varphi^2({\bf x}) + c (\nabla \varphi({\bf x}))^2 
+ u \varphi^4({\bf x}) - {\bf h} \varphi({\bf x}) \right) d {\bf x} \;,
\end{equation}
where the order parameter $\varphi({\bf x})$ is an
$n$--component vector with components $\varphi_i({\bf x})$, depending on the
coordinate ${\bf x}$, $T$ is the temperature, ${\bf h}$ is an external field.
The same model, but without the external field ${\bf h}$, has been discussed
in~\cite{K1}, representing the $\varphi^4$ term as
\begin{eqnarray}
&&\int\int \varphi^2 ({\bf x}_1) u({\bf x}_1-{\bf x}_2)
\varphi^2 ({\bf x}_2)  d{\bf x}_1 d{\bf x}_2 \label{eq:phi4} \\
&&= V^{-1} \sum\limits_{i,j,{\bf k}_1,{\bf k}_2, {\bf k}_3}
\varphi_i({\bf k}_1) \varphi_i({\bf k}_2) u_{{\bf k}_1+{\bf k}_2}
\varphi_j({\bf k}_3) \varphi_j(-{\bf k}_1 -{\bf k}_2 -{\bf k}_3)
\nonumber \;,
\end{eqnarray}
where in our special case of~(\ref{eq:H}) we have
$u({\bf x})= \delta({\bf x})$ and
$u_{\bf k} \equiv u$. This is obtained by
using the Fourier representation
$\varphi_i({\bf x}) = V^{-1/2} \sum_{k<\Lambda}
\varphi_i({\bf k}) \, e^{i{\bf kx}}$, where $V=L^d$ is the volume
of the system and $d$ is the spatial dimensionality. Like in~\cite{K1}, here
we suppose that the field $\varphi_i({\bf x})$ does not
contain the Fourier components $\varphi_i({\bf k})$ with $k>\Lambda$.
At ${\bf h}={\bf 0}$, the model undergoes the second--order phase
transition with a spontaneous long--range ordering.
Besides, all the directions of ordering are equally probable.
To remove this degeneracy, we consider the thermodynamic limit at an
infinitesimal external field, i.~e.,
$\lim\limits_{h \to 0} \lim\limits_{L \to \infty}$ where
$h=\mid {\bf h} \mid$.
In this case the magnetization ${\bf M}= \langle \varphi \rangle$
is oriented just along the external field.
We consider also a model with Hamiltonian
\begin{equation} \label{eq:H1}
H/T= \int \left( r_0 \varphi^2({\bf x}) - \delta \cdot \varphi_1^2({\bf x})
+ c (\nabla \varphi({\bf x}))^2 + u \varphi^4({\bf x})  \right) d {\bf x} \;.
\end{equation}
In the limit $\lim\limits_{\delta \to 0} \lim\limits_{L \to \infty}$
this model is equivalent to the original one at
$\lim\limits_{h \to 0} \lim\limits_{L \to \infty}$ in the sense that
the magnetization is parallel to certain axis labeled by $i=1$.
Some degeneracy still is present in~(\ref{eq:H1}), since two opposite
ordering directions are equivalent, but this peculiarity does
not play any role if we consider, e.~g., the Greens function
$\widetilde G_i({\bf x})=
\left< \varphi_i({\bf 0}) \varphi_i({\bf x}) \right>$.
In the Fourier representation, the correlation function $G_i({\bf k})$
is defined by
\begin{equation}
\left< \varphi_i({\bf k}) \, \varphi_j(-{\bf k}) \right>
= \delta_{ij} \, G_i({\bf k}) \; .
\end{equation}

Hamiltonian~(\ref{eq:H1}) is more suitable for our analysis
than~(\ref{eq:H}), since the equations derived in~\cite{K1}
can be easily generalized to include the symmetry breaking term
$\delta \cdot \varphi_1^2({\bf x})$. It is
incorporated in the Gaussian part of the Hamiltonian
\begin{equation} \label{eq:H0}
H_0/T=\sum_{i,{\bf k}} \left( r_0 - \delta \cdot \delta_{i,1} +
c {\bf k}^2 \right) \mid \varphi_i({\bf k}) \mid^2 \;.
\end{equation}
As a result, the Dyson equation in~\cite{K1} becomes
\begin{equation} \label{eq:Dyson}
\frac{1}{2G_i({\bf k})}= r_0 - \delta \cdot \delta_{i,1} +ck^2
-\frac{\partial D(G)}{\partial G_i({\bf k})} + \vartheta_i({\bf k})
\end{equation}
where $D(G)$ is a quantity, the diagram expansion of
which contains all skeleton diagrams (i.~e., those connected
diagrams without outer lines containing no parts like
\mbox{\selfe)}, constructed of the fourth order vertices \mbox{\dshv.}
The solid coupling lines in the diagrams are related to the correlation
function $G_i({\bf k})$, but the dashed lines to
$-V^{-1} u_{\bf k} = -V^{-1} \int u({\bf x}) e^{-i{\bf kx}} d{\bf x}$.
Here the notation $u_{\bf k}=u \, \widetilde u_{\bf k}$ is used
for a generalization, while the actual case of interest is
$\widetilde u_{\bf k} \equiv 1$.
Any two solid lines connected to the same kink (node) have the
same index  $i$. According to the definition, Eq.~(\ref{eq:Dyson}) is exact.
It is ensured including the remainder term $\vartheta_i({\bf k})$
which does not contribute to the perturbation
expansion in $u$ power series. Quantity $D(G)$ is given by
\begin{equation} \label{eq:D}
D(G)=D^*(G,1) + \skd
\end{equation}
where $D^*(G,\zeta)$ is the solution of the differential equation 
\begin{equation} \label{eq:Dast}
D^*(G,\zeta)=-{1 \over 2} \sum\limits_{\bf q} \ln[1-2 \Sigma({\bf q},\zeta)]
- \zeta \frac{\partial}{\partial \zeta} D^*(G,\zeta) 
\end{equation}
with the boundary condition 
\begin{equation}
D^*(G,0)=-{1 \over 2} \sum\limits_{\bf q} \ln[1-2 \Sigma^{(0)}({\bf q})] \;,
\end{equation}
where
\begin{equation} \label{eq:Sig0}
\Sigma^{(0)}({\bf q})=-2 u_{\bf q} V^{-1} \sum\limits_{i,{\bf k}}
G_i({\bf k}) G_i({\bf q}-{\bf k}) \;.
\end{equation}
Here $\Sigma({\bf q},\zeta)$ is a quantity having the diagram expansion
\begin{eqnarray} 
&& \Sigma({\bf q},\zeta) = \dg + \zeta \ddg
+ \zeta^2 \left\{ \dddg  \right. \label{eq:ee}
\\ \vspace*{2ex}
&& \left. + \dddgg + \dddggg + \dddgggg \right\} + ... \nonumber
\end{eqnarray}
including all diagrams of this kind 
which  cannot be split in two as follows \mbox{\two}. These
are skeleton diagrams with respect to both the solid and
the waved lines (i.~e., do not contain parts like \selfe and/or
\mbox{\wblock)} with factors
\begin{equation} \label{eq:wl}
\wl = -  u_{\bf k} V^{-1} \, / \, [1-2 \Sigma({\bf k},\zeta)]
\end{equation}
corresponding to the waved lines, and factor $-V^{-1} u_{\bf q}$
corresponding to a pair of fixed (formally considered as
nonequivalent) broken dashed lines in~(\ref{eq:ee}).
Quantity $\Sigma({\bf q},\zeta)$ is defined by converging sum and 
integrals (cf. Sec.~4.7 and Appendix A in~\cite{K1}),
i.~e.,
\begin{eqnarray}
\Sigma({\bf q},\zeta)&=&\Sigma^{(0)}({\bf q}) +
\int\limits_0^{u^{-p}} e^{-t_1} dt_1
\int\limits_0^{u^{-p}} e^{-t_2} B({\bf q},\zeta,t_1 t_2) \,dt_2 \;,
\label{eq:one} \\
B({\bf q},\zeta,t)&=&\sum\limits_{n=1}^{\infty}
\frac{\zeta^n t^n}{(n !)^2} \, \Sigma^{(n)}({\bf q},\zeta)
\label{eq:two} \; ,
\end{eqnarray}
where $\Sigma^{(n)}({\bf q},\zeta)$ represents the sum of diagrams of the
$n$--th order in~(\ref{eq:ee}), and $p$ is a constant $0<p<1/2$.
Note that the zeroth--order term is given by Eq.~(\ref{eq:Sig0}).

Based on these equations of the GFD theory, we have derived 
the possible values of
the exact critical exponents $\gamma$ and $\nu$ describing the divergence
of susceptibility, i.~e. $\chi \propto (T-T_c)^{-\gamma}$, and correlation
length, i.~e. $\xi \propto (T-T_c)^{-\nu}$, when approaching the critical
point $T=T_c$ from higher temperatures. These values at the spatial
dimensionality $d=2,3$ and the dimensionality of the order parameter
$n=1,2,3, \ldots$ (only the case $n=1$ is meaningful at $d=2$) are~\cite{K1}
\begin{eqnarray} 
\gamma &=& \frac{d+2j+4m}{d(1+m+j)-2j} \label{eq:resg} \\
\nu &=&   \frac{2(1+m)+j}{d(1+m+j)-2j} \label{eq:resn}
\end{eqnarray}
where $m$  may have values  $m=1, 2, 3, \ldots$, and  $j$  may have values
$j=$ $-m$, $-m+1$, $-m+2, \ldots$
A general hypothesis relating the values of $m$ and $j$ to different
models, as well as corrections to scaling for different physical
quantities and several numerical tests have been discussed in~\cite{K2,K3}.
Here we only mention that $m=3$ and $j=0$ holds at
$n=1$ to coincide with the known exact results for 2D Ising model.

Since our equations contain the diagram expansion
in terms of the true correlation function $G_i({\bf k})$ instead
of the Gaussian one, they allow an analytic continuation from
the region $r_0>0$, where they have an obvious physical solution~\cite{K1},
to arbitrary $r_0$ value. One has to start with
a finite volume, considering the thermodynamic limit afterwards.
In this paper we have extended our analysis to include the
region of negative $r_0$ values below the critical point and to
study the transverse and longitudinal fluctuations of the order
parameter field in presence of an infinitesimal external field.

\section{Alternative formulation of the basic equations}
\label{sec:alternative}

An alternative formulation of our diagrammatic equations can be helpful to
prove some basic properties of the solution (see the end of Appendix). 
Namely, a resummation of the self--energy diagrams contained in  
$R_i({\bf k})=-\partial D^*(G,1)/\partial G_i({\bf k})$ can be used instead 
of~(\ref{eq:Dast}). As discussed in~\cite{K1}, these are skeleton diagrams
like \selfek  with two outer solid lines, obtained by breaking one
solid line with wave vector ${\bf k}$ in the coupled skeleton diagrams of
$D^*(G,1)$. According to our notation, the factors of the lines
marked by crosses are omitted, and ``coupled'' means that the diagram 
does not contain outer lines. The summation over the linear chains
of blocks \ssb contained in the self--energy diagrams of $R_i({\bf k})$ can be
performed, as it has been done in~\cite{K1} with the diagrams of 
$\Sigma({\bf k},\zeta)$. It yields the expansion of 
$R_i({\bf k})$ represented by skeleton diagrams where the true
correlation function $G_i({\bf k})$ is related to the solid lines and
the dashed lines are replaced by the waved lines. Like 
in~(\ref{eq:ee}), these diagrams do not contain parts \selfe and 
\wblock. According to~(\ref{eq:D}), the diagram \skd is not included in
$D^*(G,1)$, and it also cannot be involved in the actual grouping of
diagrams since the extension of dashed line by adding the blocks \ssb
would yield non--skeleton diagrams in this case. In such a way, we
have the expansion
\begin{equation} \label{eq:altern}
-R_i({\bf k}) = \skedd  + \skeddd + \ldots
\end{equation}
Here we have indicated explicitly that certain index $1 \le i \le n$
refers to the outer lines in the $n$--component case. 
The waved line, given by~(\ref{eq:wl}), corresponds to $\zeta=1$.
The combinatorial factors 4, 32, etc., are included in the diagrams
not distinguishing the two orientations with respect to the vectors
${\bf k}$ and $-{\bf k}$ as different. The resummation of  
expansion~(\ref{eq:altern}) can be made by adopting our method 
presented by Eqs.~(\ref{eq:one}) and~(\ref{eq:two}).

The Dyson equation~(\ref{eq:Dyson}) can be formulated in accordance with 
the variational principle, i.~e., it follows from the extremum condition
$\partial \widetilde F(G) / \partial G_i({\bf k}) = 0$ of the 
reduced free energy function
\begin{equation}
\widetilde F(G)= -\frac{1}{2} \sum\limits_{i,{\bf k}} 
\ln \left[ 2 \pi G_i({\bf k}) \right]
+\sum\limits_{i,{\bf k}} \left[ \theta_i({\bf k}) G_i({\bf k})
- \frac{1}{2} \right] \, - \, D(G) \;.
\end{equation}
Here $\theta_i({\bf k})=r_0-\delta \cdot \delta_{i,1}+ck^2$ and
$\widetilde F(G)$ depends on the set of discrete variables
$G_i({\bf k})$, as consistent with the diagrammatic definition of $D(G)$.
This function is related to the free energy $F=-T \ln Z$
via $\widetilde F(G)=F/T$. First it has been obtained in~\cite{MK} for the
case $n=1$. The fact that $\widetilde F(G)$ provides
the diagrammatic representation of the reduced free energy can be proven as 
follows. According to~(\ref{eq:Dyson}), 
$\partial \widetilde F(G) / \partial G_i({\bf k}) = 0$
holds, neglecting the remainder term $\vartheta_i({\bf k})$. It means that 
\begin{equation} \label{eq:derf}
\frac{\partial \widetilde F}{\partial r_0} =
\sum\limits_{i,{\bf k}} G_i({\bf k}) = V \langle \varphi^2({\bf x}) \rangle
\equiv \frac{\partial (F/T)}{\partial r_0}
\end{equation}
is true within the diagrammatic perturbation theory. Besides, it is 
straightforward to check that $\widetilde F(G)=F/T$
holds at $r_0 \to + \infty$. By integration in~(\ref{eq:derf}) over
$r_0$ from $+\infty$ to any finite value we find that $\widetilde F(G)=F/T$
is valid in general.

\section{The correlation function and susceptibility below $T_c$}
\label{sec:below}

    Some important relations between the correlation function,
the long--range order parameter $M$ (e.~g., magnetization or polarization),
and susceptibility $\chi$, following
directly from the first principles, are considered in this section.

We have defined the correlation function in the coordinate representation as
\begin{equation}
\widetilde G_i({\bf x}) = \left< \varphi_i({\bf x}_1) \,
\varphi_i({\bf x}_1+{\bf x}) \right>
= V^{-1} \sum\limits_{\bf k} G_i({\bf k}) \, e^{i{\bf kx}} \; .
\end{equation}
For simplicity, first, let us consider the one--component case. In this case
(omitting the index $i$) we have~\cite{Baxter}
\begin{equation} \label{eq:M}
M^2 = \lim\limits_{x \to \infty} \widetilde G({\bf x})
= G({\bf 0})/V \hspace{3ex} \mbox{at} \hspace{2ex} V \to \infty \; ,
\end{equation}
or
\begin{equation} \label{eq:Gx}
\widetilde G({\bf x}) = M^2 + \widetilde G'({\bf x}) \; ,
\end{equation}
where $\widetilde G'({\bf x})$ tends to zero if $x \to \infty$. In the
Fourier representation~(\ref{eq:Gx}) reduces to
\begin{equation} \label{eq:Gbel}
G({\bf k}) = \delta_{{\bf k},{\bf 0}} \, V M^2 + G'({\bf k})
\end{equation}
where $G'({\bf k})$ is the Fourier transform of $\widetilde G'({\bf x})$.
The susceptibility, calculated directly from the Gibbs distribution, is
\begin{equation} \label{eq:chi}
\chi = \lim\limits_{h \to 0} \lim\limits_{L \to \infty}
\frac{\partial \langle \varphi \rangle}{\partial h}
= \lim\limits_{h \to 0} \lim\limits_{L \to \infty}
\int \left( \widetilde G({\bf x}) - {\langle \varphi
\rangle}^2 \right) \, d{\bf x} = G'({\bf 0}) \; .
\end{equation}
In this limit $\langle \varphi \rangle = M$ holds, the latter,
however, is not correct at $h=0$ when $\langle \varphi \rangle = 0$.
The considered limit for $\chi$ exists and
$G'({\bf 0})$ has a finite value in our case
of $n=1$, since the correlation function $\widetilde G'({\bf x})$ is
characterized by a finite correlation length $\xi$, which ensures the
convergence of the integral in~(\ref{eq:chi}).

  Consider now the case $n>1$. If an external field is applied along
the $i$--th axis with $i=1$ (even if $h \to 0$), the longitudinal
Greens function $G_{\parallel}({\bf k}) \equiv G_1({\bf k})$ 
behaves in a different way than the transverse one
$G_{\perp}({\bf k}) \equiv G_j({\bf k})$ with $j \ne 1$. It is
a rigorously stated fact~\cite{Ma} that
$G_{\perp}({\bf 0})$ diverges as $M/h$  if $h \to 0$  below $T_c$, which
is related to the divergence of the transverse susceptibility in this case.
In analogy to~(\ref{eq:Gbel}) and~(\ref{eq:chi}) we have
\begin{eqnarray}
G_{\parallel}({\bf k}) &=& \delta_{{\bf k},{\bf 0}} \, V M^2 +
G'_{\parallel}({\bf k}) \;, \label{eq:Gbelp} \\
\chi_{\parallel}(h) &=& \partial M / \partial h =
G'_{\parallel}({\bf 0}) \;. \label{eq:longsus}
\end{eqnarray}
Our further analysis shows that the longitudinal susceptibility
$\chi_{\parallel}(h)$ diverges at $h \to 0$ for $2<d<4$ and $n>1$, i.~e.,
$G'_{\parallel}({\bf k})$ diverges at $k \to 0$. Note that
$G'_{\parallel}({\bf k}) = G_{\parallel}({\bf k})$ holds at
${\bf k} \ne {\bf 0}$. The long-wavelength divergences of the transverse and
longitudinal correlation functions below $T_c$ is known in literature
as the Goldstone mode singularities established by the Goldstone
theorem~\cite{Goldstone,Wagner}.

\section{Generalized scaling hypothesis}
\label{sec:gsh}

According to the known~\cite{Ma} scaling hypothesis, the correlation function
above the critical point, i.~e. at $T >T_c$ and $T \to T_c$, can be
represented in a scaled form
\begin{equation} \label{eq:sc1}
G_i({\bf k}) \simeq \xi^{2-\eta} g_i(k \xi) \;,
\end{equation}
where $\xi$ is the correlation length, $\eta$ is the critical exponent,
and $g_i(z)$ is a scaling function. Since $\xi \sim t^{-\nu}$ holds,
where $t=(T/T_c)-1$ is the reduced temperature, the correlation function
can be represented also as
\begin{equation} \label{eq:sc2}
G_i({\bf k}) \simeq t^{-\gamma} g_i \left( k t^{-\nu} \right) \;,
\end{equation}
where $\gamma=(2-\eta) \nu$. Since the phase transition occurs
merely at a single point $h=t=0$ in the $h$--$t$ plane, there exists
a way how the scaling relations like~(\ref{eq:sc1}) or~(\ref{eq:sc2})
can be continued to the region $t<0$ passing the singular point $h=t=0$.
Eq.~(\ref{eq:sc1}) is not valid at $h=0$ and $t<0$ in the case of $n>1$,
since $G'_{\parallel}({\bf 0})$ and, therefore, the correlation length $\xi$
diverges at $h \to 0$ [cf. Eq.~(\ref{eq:chi})]. 
The known scaling relations are recovered assuming that
the physical picture remains
similar if we approach the critical point like $t \to s \, t$ and
$h \to s^{\sigma} h$, where $s<1$ is the rescaling factor. 
Thus, the distance from the critical point 
$\hat t=\left( t^2 + h^{2/\sigma} \right)^{1/2}$
and the polar angle $\theta= \arctan \left( h^{1/\sigma}/t \right)$ in the
$t$ -- $h^{1/\sigma}$ plane are two relevant scaling arguments.
According to this discussion, a suitable generalization
of the scaling relation~(\ref{eq:sc2}) is
\begin{equation} \label{eq:scg}
G_i({\bf k}) \simeq {\hat t}^{-\gamma} g_i \left( k {\hat t}^{-\nu},
\theta \right) \;,
\end{equation}
which is true at $\hat t \to 0$ for any given values of $k {\hat t}^{-\nu}$
and $\theta$. Consider $G_{\perp}({\bf 0})$ at a small negative $t$.
Taking into account that $h^{1/\sigma} \simeq \hat t \, (\pi -\theta)$ and
$M \propto (-t)^{\beta} \simeq h^{\beta/\sigma} (\pi -\theta)^{-\beta}$
hold at $\theta \to \pi$, the correct result
$G_{\perp}({\bf 0}) \simeq M/h$ is obtained in this limit
if  $g_{\perp} (0,\theta) \propto (\pi -\theta)^{-\sigma}$ holds
at $\pi -\theta \to 0$ and the scaling dimension is
$\sigma= \beta + \gamma$. By generalizing the scaling
relation $M \propto (-t)^{\beta}$ to $M \propto {\hat t}^{\beta}$
(at a fixed $\theta$) we obtain also the correct behavior
$M \propto h^{1/\delta}$ at $t=0$, where $\delta=1+\gamma/\beta$.
Eq.~(\ref{eq:scg}) makes sense for $G_{\parallel}({\bf k})$ at
${\bf k} \ne {\bf 0}$. According to~(\ref{eq:Gbelp}) and~(\ref{eq:longsus}),
the longitudinal susceptibility is
$\chi_{\parallel}=G'_{\parallel}({\bf 0})=G_{\parallel}(+0)$,
where $G_{\parallel}(+0)$ denotes the value of the Greens function
at an infinitely small, but nonzero $k$ value. It is easy to check
that~(\ref{eq:scg}) reproduces the known scaling behavior of
$\chi_{\parallel}$ for $t \ge 0$ both at $h=0$
($\chi_{\parallel} \propto t^{-\gamma}$) and at $t=0$ 
($\chi_{\parallel} \propto h^{\frac{1}{\delta}-1}$).
 In the limit $h \to 0$ Eq.~(\ref{eq:scg}) yields
\begin{eqnarray}
G_i({\bf k}) &\simeq& \mid t \mid^{-\gamma}
g_i^+ \left( k \mid t \mid^{-\nu} \right) \hspace{4ex} \mbox{at}
\hspace{2ex} t>0 \label{eq:scp} \\
G_i({\bf k}) &\simeq& \mid t \mid^{-\gamma}
g_i^- \left( k \mid t \mid^{-\nu} \right) \hspace{4ex} \mbox{at}
\hspace{2ex} t<0 \label{eq:scm} \;,
\end{eqnarray}
where
$g_i^+ (z) = g_i (z, 0)$ and $g_i^- (z) = g_i (z, \pi)$.
The analysis of our diagrammatic equations confirm
the scaling relations~(\ref{eq:scp}) and~(\ref{eq:scm}). It shows
also that, in the case of the order parameter dimensionality $n>1$,
both the longitudinal and the transverse Greens functions diverge
at $k \to 0$ when $T<T_c$. It means that 
$g_i^-(z)$ diverges at $z \to 0$ for $n>1$. In any case
we have $g_i^{\pm}(z) \propto z^{-2+\eta}$ at $z \to \infty$,
which means that the correlation function continuously transforms
to the known critical Greens function $G_i({\bf k}) \sim k^{-2+\eta}$ at
$t \to 0$.

\section{Exact scaling relations and their consequences} \label{sec:exact}

In distinction to Sec.~\ref{sec:gsh}, here we consider other
kind of scaling relations which also are relevant to our further
analysis. These are exact and rigorous relations between the
correlation function and parameters of the Hamiltonian
$H/T = H_0/T + H_1/T$, where $H_0/T$ is given by~(\ref{eq:H0})
and $H_1/T$ represents the $\varphi^4$
contribution~(\ref{eq:phi4}) at $u_{\bf k}=u$.

Following the method described in Appendix~B of~\cite{K1},
the Hamiltonian $H/T$ is transformed to


\begin{eqnarray} 
H/T &=& \sum_{i,{\bf p}} \left( R - \epsilon \cdot \delta_{i,1} +
 {\bf p}^2 \right) \mid \Psi_i({\bf p}) \mid^2 \label{eq:Htr} \\
&+& V_1^{-1} \sum\limits_{i,j,{\bf p}_1,{\bf p}_2, {\bf p}_3}
\Psi_i({\bf p}_1) \Psi_i({\bf p}_2) 
\Psi_j({\bf p}_3) \Psi_j(-{\bf p}_1 -{\bf p}_2 -{\bf p}_3) \nonumber \;,
\end{eqnarray}
where $\Psi_i({\bf p})=u^{\alpha} c^{-d \alpha/2} \varphi_i({\bf k})$,
${\bf p}=c^{2 \alpha} u^{-\alpha} {\bf k}$,
$R=r_0 \, c^{d \alpha} u^{-2 \alpha}$,
and $\epsilon=\delta u^{-2 \alpha} c^{d \alpha}$. Here $\alpha =1/(4-d)$
and the summation over ${\bf p}$ takes place within
$p<p_0=c^{2 \alpha} u^{-\alpha} \Lambda$ in accordance with the rescaled
volume $V_1=V c^{-2d \alpha} u^{d \alpha}$.
According to~(\ref{eq:Htr}) we have
\begin{equation}
G_i({\bf k}) = \left< \mid \varphi_i({\bf k}) \mid^2 \right>
=c^{d \alpha} u^{-2 \alpha} \left< \mid \Psi_i({\bf p}) \mid^2 \right>
=c^{d \alpha} u^{-2 \alpha} \widetilde g_i({\bf p},p_0,R,\epsilon,V_1) \;,
\end{equation}
where, at fixed $d$ and $n$, $\widetilde g_i$ is a function of the given
arguments only. The thermodynamic limit exists at ${\bf k} \ne {\bf 0}$, so
that in this case we can write
\begin{equation} \label{eq:s}
\lim\limits_{\delta \to 0} \lim\limits_{V \to \infty} G_i({\bf k})
=c^{d \alpha} u^{-2 \alpha} \hat g_i({\bf p},p_0,R) \;,
\end{equation}
where the scaling function $\hat g_i$ represents the limit of
$\widetilde g_i$. If $R$ is varied, then the model with
Hamiltonian~(\ref{eq:Htr}) undergoes the second--order phase
transition at some critical value
$R=R_c(p_0)<0$.
 Thus, Eq.~(\ref{eq:s})
can be rewritten in new variables with a scaling function
$g_i({\bf p},p_0,t)$, where
$t=1-(R/R_c)=1-(r_0/r_{0c})$ is
the reduced temperature, $r_{0c}$ being the critical value of $r_0$.
Thus, we have
\begin{equation} \label{eq:ss}
\hat g_i({\bf p},p_0,R) = g_i({\bf p},p_0,t) \;.
\end{equation}
Based on~(\ref{eq:s}) and~(\ref{eq:ss}), we can make some conclusions
about the scaling of the critical region for the reduced temperature $t$,
as well as for the wave vector ${\bf k}$ at $t=0$ and also
at a fixed $t<0$. The latter case is relevant for the
long--wave limit $k \to 0$ at $n>1$. By the critical region we mean
the region inside of which the correlation function is well described
by a certain asymptotical law. According to~(\ref{eq:s}) and~(\ref{eq:ss}),
the width of the critical region $t_{crit}$ or $k_{crit}$,
as well as the coefficients in asymptotic expansions in powers of $k$ 
at a fixed $t$ ($t=0$ for $n \ge 1$ or $t<0$ for $n>1$) can be written
in a scaled form with (or without) power--like prefactors
and scaling functions containing single argument
$p_0=c^{2 \alpha} u^{-\alpha} \Lambda$.

The limit $u \to 0$ is important in our consideration. To
extract exact critical exponents from our equations, it has to be
ensured that, inside the critical region, the remainder
term $\vartheta_i({\bf k})$ in~(\ref{eq:Dyson}) is much smaller than any term
in the asymptotic expansion of $1/G_i({\bf k})$. This is possible at
$u \to 0$ if these expansion terms and also the width of the critical region
do not tend to zero faster than any positive power of $u$~\cite{K1}.
For the scaling functions of $p_0$,
the limit $u \to 0$
is equivalent to the limit $\Lambda \to \infty$ at $d<4$.
The relevant quantities
($t_{crit}$, $k_{crit}$, and expansion coefficients at $k$ powers)
can tend to zero exponentially at $u \to 0$ only if the corresponding
scaling functions of $c^{2 \alpha} u^{-\alpha} \Lambda$ do so.
The latter would mean
that these quantities are very strongly (exponentially) affected
by any relatively small variation of the upper cutoff parameter
$\Lambda$ at large $\Lambda$ values. It seems to be rather unphysical,
since the long--wave behavior at a fixed $t$ (also at $t<0$) cannot be
so sensitive to small
variations in the short--range interactions. Due to the
joining of the asymptotic solutions, the fact that the expansion
coefficients in $k$ power series at $t=0$ are not
exponentially small in $u$
means also that the same is true for the expansion coefficients 
in $\mid t \mid$ power series at $\mid t \mid \to 0$. Thus,
only the remainder term $\vartheta_i({\bf k})$ tends to zero
faster than $u^s$ at any $s>0$, provided that our solution
represents an analytic continuation (see the end of Sec.~\ref{sec:pri})
from the stable domain $r_0>0$ where our equations have originated and
where this basic property of $\vartheta_i({\bf k})$ follows directly from
our derivations.
It implies, in particular, that the solution below $T_c$ should coincide
with~(\ref{eq:scm}), as consistent with the existence of continuous
second--order phase transition.
If we have a smooth solution of this kind, then the critical exponents
can be determined by considering suitable limits~\cite{K1}
($u \to 0$ and $k \sim u^{r} k_{crit}(u)$,
or $u \to 0$ and $t \sim u^{r/\nu} t_{crit}(u)$ with $r>0$)
not only at $T=T_c$ and $T \to T_c$, but also at $T<T_c$.
In this case the remainder term $\vartheta_i({\bf k})$
is negligibly small~\cite{K1}.

\section{The low--temperature solution at $n=1$} \label{sec:low1}

Let us now consider the solution of our equations below
the critical point starting with the case $n=1$.
The symmetry breaking term with $\delta$ is irrelevant at $n=1$,
therefore we set $\delta=0$.
According to~(\ref{eq:M}), $1/G({\bf k})$ vanishes at ${\bf k}={\bf 0}$
in the thermodynamic limit $V \to \infty$,
so that the equation~(\ref{eq:Dyson}) (taking into account~(\ref{eq:D})
and omitting the irrelevant correction $\vartheta_i({\bf k})$) for
scalar order--parameter field ($n=1$) can be written as
\begin{eqnarray} 
1 / (2 G_1({\bf k})) &=& ck^2 
+ R_1({\bf k}) - R_1({\bf 0}) \hspace{4ex} \mbox{at} \hspace{2ex}
{\bf k} \ne {\bf 0} \label{eq:G} \; ,  \\
1 / (2 G_1({\bf 0})) &=& r_0 +2u \widetilde G
+ R_1({\bf 0}) = 0 \label{eq:condi} \; ,
\end{eqnarray}
where (for arbitrary $n$)
\begin{equation} \label{eq:tilG}
\widetilde G = \left< \varphi^2({\bf x}) \right>
= V^{-1} \sum\limits_{i,{\bf k}} G_i({\bf k}) \; ,
\end{equation}
\begin{equation} \label{eq:R}
R_i({\bf k}) =
- \frac{\partial D^*(G,1)}{\partial G_i({\bf k}) } \; .
\end{equation}
To simplify the notation, further we shall omit the index $i \equiv 1$
in the actual case of $n=1$.

According to~(\ref{eq:M}), single term with ${\bf k}={\bf 0}$ gives
a nonvanishing contribution to~(\ref{eq:tilG}) at $V \to \infty$, while
the contribution of all other terms may be replaced by an integral, i.~e.,
\begin{equation} \label{eq:wG}
\widetilde G = M^2 + (2 \pi)^{-d} \int G'({\bf k}) \, d {\bf k} \; .
\end{equation}
Similarly, terms with $M^2$, $M^4$, $M^6$, etc. appear in~(\ref{eq:ee})
due to the contributions provided by zero--vectors related to
some of the solid lines. For instance, the zeroth--order term~(\ref{eq:Sig0})
reads
\begin{equation} \label{eq:Sig00}
\Sigma^{(0)}({\bf q})=-2 u
\left( 2M^2 \, G({\bf q})+ (2 \pi)^{-d}
\int G'({\bf k}) G'({\bf q}-{\bf k}) d{\bf k} \right) \;.
\end{equation}
The terms with the spontaneous magnetization $M$ appear in our equations
in a natural way if we first consider a very large, but finite
volume $V$, which then is tended to infinity. They appear as a feedback
which does not allow the right hand side of~(\ref{eq:condi}) to become
negative, by keeping it at $1/(2G({\bf 0})) \sim 1/V$, when $r_0$ goes
to large enough negative values.

The actual model at $n=1$ belongs to the Ising universality class
characterized by a finite correlation length at $T<T_c$. It means that
$G'({\bf 0})$ has a finite value and $1/G({\bf k})$
transforms to zero at ${\bf k}={\bf 0}$ by a jump. According
to~(\ref{eq:G}), it means that $R(+0) \ne R({\bf 0})$ holds,
where the value of $R({\bf k})$ at an infinitesimal non--zero $k$ is denoted
by $R(+0)$. To show that this is really possible, consider the contribution
(denoted by $R^{(0)}({\bf k})$) of the first diagram in~(\ref{eq:ee})
which yields
\begin{equation} \label{eq:Rp0}
R^{(0)}(+0)=R^{(0)}({\bf 0}) + \frac{4u \, M^2}{1+4u \left(
2M^2 \, G'({\bf 0}) + (2 \pi)^{-d} \int {G'}^2({\bf q}) \, d{\bf q}
\right)} \; .
\end{equation}
From~(\ref{eq:Rp0}) we see that $R(+0) \ne R({\bf 0})$ holds, in general,
if $G'({\bf 0})$ has a finite value.
Therefore such a selfconsistent solution is, in principle, possible.

  Consider now the solution at $r_0 \to -\infty$ and small $u$, i.~e.,
at low temperatures. In this case we find a solution such that
\begin{equation} \label{eq:sob}
M^2 = -r_0/(2u) \;, \hspace{3ex} G({\bf k})= -1/ \left(A \, r_0 \right) \;,
\hspace{3ex} \Sigma({\bf k},\zeta)= f(\zeta,A) \;, 
\end{equation}
hold at any fixed $u>0$ and ${\bf k} \ne {\bf 0}$, if $r_0 \to -\infty$,
where $A$ is a constant independent of $r_0$, 
$c$, and $\Lambda$, $f(\zeta,A)$ is a function of the given arguments.
One expects that $A$ tends to some universal constant at $u \to 0$.
In this case, at $A=4$, our solution coincides with the Gaussian
approximation
\begin{equation} \label{eq:Gausappr}
G'({\bf k}) = \frac{1}{-4r_0 + 2 ck^2} \; .
\end{equation}
The correction $\sim ck^2$ has been neglected in~(\ref{eq:sob}).

Condition $M^2=-r_0/(2u)$, corresponding to the
minimum of~(\ref{eq:H1}), holds for any physical solution if $r_0 \to -\infty$
since fluctuations are suppressed. This follows
from~(\ref{eq:condi}) and~(\ref{eq:wG}), if the main terms
$r_0$ and $2u \, M^2$ are retained in~(\ref{eq:condi}).

Quantity $\Sigma({\bf 0},\zeta)$ in the denominator of~(\ref{eq:wl})
diverges in the thermodynamic limit\linebreak
$V \to \infty$ like $V$ or even faster, i.~e., like $V^{\mu}$ with
$\mu \ge 1$, as shown in Appendix.
The divergence appears due to the special contributions provided by
zero wave vectors ${\bf k}={\bf 0}$ related to some of the solid lines
in the diagram expansion~(\ref{eq:ee}) of $\Sigma({\bf 0},\zeta)$ at $T<T_c$.
These diverging terms make the analysis below $T_c$ more difficult
as compared to the case $T \ge T_c$ discussed in~\cite{K1}.

In spite of the divergence of $\Sigma({\bf 0},\zeta)$, a single term with
${\bf q}={\bf 0}$ in~(\ref{eq:Dast}) does not contribute to
$D^*(G,\zeta)$ and $R({\bf k})$ if $V \to \infty$. Really, if
$\Sigma({\bf 0},\zeta)$ diverges as $- \left( V/V_0 \right)^s$
($V_0$ is the volume of an elementary cell) with $s>0$, then this single
term is approximately $-(s/2) \cdot \ln \left( V/V_0 \right)$, whereas the
whole sum is proportional to $V/V_0$. On the other
hand, $\Sigma({\bf 0},\zeta)$  appears in the denominator of the corresponding
term if the derivative with respect to $G({\bf k})$ is calculated
in~(\ref{eq:Dast}), therefore, this term cannot be by a factor
$V$ larger than other terms, i.~e., it cannot give an especial contribution.

The perturbation sum for $\Sigma({\bf k},\zeta)$ at 
${\bf k} \ne {\bf 0}$ also contains terms diverging at $V \to \infty$. 
In a normal case,
the constraint ${\bf k} = {\bf 0}$ for wave vectors of $m$ solid lines
in a diagram means removal of $m$ integrations over wave vectors.
However, for some distributions of the zero--vectors this condition is
violated, which yields the diverging terms.
We have shown in Appendix that the divergent
terms contain insertions like, e.~g., \fourz  with $2m$ outer
solid lines having fixed ${\bf k}={\bf 0}$ vectors. 
As shown in Appendix, a resummation of these insertions
gives a non--divergent result. Besides, the simplified analysis
which ignores these insertions is qualitatively correct, as regards the
general scaling form of the solution.
However, for a complete formal correctness we should take into account the 
fact that specific values of scaling functions can be renormalized
by these zero--vector--cumulant (see Appendix for explanation) insertions. 
Further we shall call the terms without such insertions the ``normal'' 
terms or contributions. Our results at $T<T_c$ are completely consistent
with those at $T=T_c$ and $T>T_c$ and agree with the non--perturbative RG
analysis provided in Sec.~\ref{sec:rg}.
It confirms the statements made in this paragraph.

An important property of the ``normal'' contributions to 
$\Sigma({\bf k},\zeta)$ at ${\bf k} \ne {\bf 0}$ is that any term, 
where zero wave vector is related to a 
waved line, vanishes in the thermodynamic limit $V \to \infty$.
It holds because the waved line vanishes due to the 
divergence of $\Sigma({\bf 0},\zeta)$.

Condition $\Sigma({\bf k},\zeta)= f(\zeta,A)$ at ${\bf k} \ne {\bf 0}$ holds
if $r_0 \to -\infty$ because $\Sigma^{(0)}({\bf q})$ and all terms of the
sum~(\ref{eq:two}) in this limit depend merely on parameters $\zeta$
and $A$ for any fixed $u$. The latter is true in the approximation
where no 
zero--vector--cumulants
are included, since
the main terms come from the diagrams in~(\ref{eq:ee}) if we extract the
contribution, containing no integrals,
where one half of the solid lines have zero wave vectors ${\bf k}={\bf 0}$
(it yields a factor $M^2$ for the first
diagram, $M^4$ -- for the second diagram, and so on).
Besides, the waved lines should have nonzero wave vectors, as explained
before. According to~(\ref{eq:sob}) each replacement of
$G({\bf 0})/V=M^2$ by an integral over wave vectors produces a factor
$\sim r_0^{-2}$. This is the reason why the main terms at
$r_0 \to -\infty$ contain no integrals.

 These main terms lead to asymptotic solutions (at $r_0 \to -\infty$)
for quantities\linebreak
$\partial \Sigma({\bf q},\zeta) / \partial G({\bf k})$
and $\partial D^*(G,\zeta) / \partial G({\bf k})$ at
${\bf q}={\bf k} \ne {\bf 0}$
in the form where $r_0$ is multiplied by some function of $\zeta$ and
$A$, namely, $R({\bf k})= - Q(A) \, r_0$.
This main contribution
vanishes at ${\bf k}={\bf 0}$, since
the derivation $\partial \Sigma({\bf 0},\zeta) / \partial G({\bf 0})$,
implicated in the calculation of $R({\bf 0})$,
means removal of a diverging factor $G({\bf 0}) \sim V$.
Quantity $\Sigma({\bf 0},\zeta)$ diverges, whereas
the waved lines in this case contain
vanishing factors $\sim 1/\Sigma({\bf 0},\zeta)$ providing
finite value of the derivative at $V \to \infty$.
Therefore, by substituting
$R({\bf k})= - Q(A) \, r_0$
into~(\ref{eq:G}) and retaining only
the leading terms, we obtain a selfconsistent equation
$2 Q(A)=A$ for the unknown amplitude $A$. Following the consideration
in Appendix, the inclusion of zero--vector--cumulant insertions
can merely renormalize the function $Q(A)$.

\section{Asymptotic solution at $T \to T_c$}  \label{sec:Asbel}

Our equations provide the asymptotic solution at $T \to T_c$,
where $T<T_c$, in the scaled form~(\ref{eq:scm}) which allows a unified
description provided here for $n \ge 1$.

Like in~\cite{K1}, here we assume that $r_0$ is the only parameter
in~(\ref{eq:H1}) which depends on temperature $T$ and the dependence
is linear. At the critical point $r_0=r_{0c}$ we have $1/G_i(+0)=0$
for all $i$ at
$\lim\limits_{\delta \to 0} \lim\limits_{V \to \infty}$, so
that the Dyson equation~(\ref{eq:Dyson}) in this limit reads
\begin{eqnarray}
&&\frac{1}{2G_i(+0)} = -\frac{dr_0}{dT} \cdot \Delta
+2u \, \left( \widetilde G - \widetilde G^* \right)
+R_i(+0)-R_i^*(+0) \label{eq:viens} \\
&&\frac{1}{2G_i({\bf k})}-\frac{1}{2G_i(+0)} =
ck^2 +R_i({\bf k})-R_i(+0) \label{eq:divi} \;.
\end{eqnarray}
Here $\Delta= \mid T-T_c \mid$ is an analog of $\mid t \mid$
in~(\ref{eq:scm}), $\widetilde G^*$ and $R_i^*(+0) \equiv R_i^*({\bf 0})$
are the values of $\widetilde G$ and $R_i({\bf 0})$ calculated at the
critical Greens function $G_i({\bf k})=G_i^*({\bf k})$ considered as a
known fixed quantity. Due to the symmetry breaking
term $\delta \cdot \delta_{i,1}$ in~(\ref{eq:H1}), only the
longitudinal component $1/G_1({\bf 0}) \equiv 1/G_{\parallel}({\bf 0})$
becomes as small as $\sim 1/V$ when $r_0$ is decreased below $r_{0c}$,
giving rise to the magnetization $M^2=G_{\parallel}({\bf 0})/V$.
According to~(\ref{eq:divi}), the condition
$1/G_{\parallel}({\bf 0})=0$ at $V \to \infty$ means
\begin{equation} \label{eq:tris}
1/ \left( 2G_{\parallel}(+0) \right) = R_{\parallel}(+0)
-R_{\parallel}({\bf 0}) \;.
\end{equation}

Equations~(\ref{eq:viens}) and~(\ref{eq:divi}) are analogous
to (48) and (49) in~\cite{K1} derived for $T>T_c$, and similar
method of analysis is valid. Namely, correct results for the
Greens function within the asymptotical region $k \sim 1/\hat \xi$ can
be obtained considering the limit $u \to 0$ and
$\Delta \sim u^{r/\nu} \Delta_{crit}(u)$, and formally
cutting the integration over $G_i({\bf k})$ and $G_i^*({\bf k})$
in~(\ref{eq:viens})
and~(\ref{eq:divi}) by $k<\Lambda= u^{-r} k_{crit}^{-1}(u)/\hat \xi$.
Here $\Delta_{crit}(u)$ and $k_{crit}(u)$ are the widths of
the critical regions for $\Delta$ and $k$, respectively, $r$ is any positive
constant, and $\hat \xi$ is an analog of the correlation length.
According to the generalized scaling hypothesis in Sec.~\ref{sec:gsh},
one may set $\hat \xi = \xi(T=T+\Delta)$.
Note that our equations~(\ref{eq:viens}) to~(\ref{eq:tris})
do not contain $r_{0c}$. Only such a form is acceptable
in this analysis: contrary to the critical exponents
the critical temperature is essentially affected
by the short--wave fluctuations.

We start our analysis with the ``normal'' (not diverging at $V \to \infty$)
terms discussed in Sec.~\ref{sec:low1} by using the well known scaling relation 
\begin{equation} \label{eq:scbeta}
2 \beta=(d-2+\eta) \nu=d \nu -\gamma
\end{equation}
relevant to $M \sim \Delta^{\beta}$.
Our diagram equations become exact in the theoretical
limits considered, and it allows in principle to find the exact critical
exponents by solving these equations (Sec.~\ref{sec:exact}). 
Therefore, the asymptotic solution
satisfies the existing exact relations between the critical
exponents, including (48). In fact, (\ref{eq:scbeta}) is the necessary condition 
at which the diagrams below $T_c$ have certain scaling properties, similar to those 
above $T_c$, where the exponents are common for the diagrams of all orders.

Selfconsistent solutions for $\Sigma({\bf q},\zeta)$ and for
$\partial D^*(G,\zeta) / \partial G_i({\bf k})$  can be found
at $2<d<4$, having similar form as in the case $T>T_c$~\cite{K1}.
This is true for the ``normal'' terms
because partial contributions in~(\ref{eq:one}) proportional
to $M^0$, $M^2$, $M^4$, etc., (corresponding to cases where zero--vectors
are related to 0, 1, 2, etc., solid lines in diagrams~(\ref{eq:ee})) have the
same form and the same common factor $\Delta^{-2 \gamma+d \nu}$ as
at $T>T_c$. It is easy to prove this property for diagrams of all orders by
a simple normalization of all wave vectors to
$\Delta^{\nu}$, like ${\bf q}'={\bf q} / \Delta^{\nu}$,
with the cutting of integration discussed above.
Contrary to the case $T>T_c$, the main term of
$\partial \Sigma({\bf q},\zeta) / \partial G_i({\bf k})$ at $T<T_c$
for given $u$, $d$, and $n$ has the form
\begin{equation} \label{eq:Sigbel}
\partial \Sigma({\bf q},\zeta) / \partial G_i({\bf k})
= V^{-1} \Delta^{-\gamma} \, Y_i({\bf q}',{\bf k}',\zeta)
+ \delta_{i,1} \delta_{{\bf k},{\bf q}} \, \Delta^{d \nu -\gamma} \,
\hat Y({\bf q}',{\bf k}',\zeta) 
\end{equation}
obtained by the above normalization, where ${\bf k}'={\bf k} \Delta^{-\nu}$. 
The additional (second) term is due to the following. If in some diagrams
of~(\ref{eq:ee}) the wave vectors for some solid lines are fixed
${\bf k}_1={\bf 0}$ (this yields an especial contribution merely
at $T<T_c$), then the wave vectors of some other solid lines also have a
fixed value ${\bf k}_2={\bf q}$. This produces the second (extraordinary)
term after derivation of the lines with a fixed wave vector ${\bf q}$.
The simplest example is provided by the first diagram in~(\ref{eq:ee})
represented by~(\ref{eq:Sig00}). The first term
in~(\ref{eq:Sig00}), which at arbitrary $n$ reads
$-4u \, M^2 G_1({\bf q})$, 
 yields an extraordinary contribution
$-4u \, \delta_{i,1} \delta_{{\bf q},{\bf k}} M^2$ to~(\ref{eq:Sigbel}).
In general, any diagram of~(\ref{eq:ee}) can be represented
as two separate parts connected by $m$ solid lines, e.~g.,
 like \mbox{\blocks.}
Each such representation gives an extraordinary contribution
provided that no more than one pair of the connecting lines
meet at the same node like \mbox{\wlpart.}
Since solid lines always make closed loops,
$m$ is even number. The same index $i$ is related to all lines of one
loop, therefore, even number of lines among those connecting
the blocks can be associated with $i=1$. Thus, an extraordinary
contribution is provided by fixing $i=1$ and ${\bf k}_1={\bf 0}$ for
$m-1$ connecting lines, since this implies the constraint $i=1$ and
${\bf k}_2={\bf q}$ for the remaining one line.
However, only those configurations give a nonvanishing contribution
at $V \to \infty$ where no more than one wave vector ${\bf k}_1={\bf 0}$
refers to each of the configurations \mbox{\wlpart.} In the opposite case
the waved line also has zero wave vector
yielding vanishing contribution
due to the divergence of $\Sigma({\bf 0},\zeta)$.

The additional term in~(\ref{eq:Sigbel}) essentially differs from the
ordinary (first) term by factor $\delta_{{\bf k},{\bf q}} V \Delta^{d \nu}$
because in this case the derivation procedure does not mean removal of
integration over wave vectors. Both terms provide
$\partial D^*(G,\zeta) / \partial G_1({\bf k}) = \Delta^{\gamma} \,
\widetilde r({\bf k}',\zeta)$, single term with ${\bf q}={\bf 0}$
in~(\ref{eq:Dast}) being negligible, as discussed in~Sec.\ref{sec:low1}.
This scaled form is valid also for $i>1$ and agree with that at $T>T_c$
derived in~\cite{K1}.
The term with $\delta_{{\bf k},{\bf q}}$ does not give a contribution to
$\partial D^*(G,\zeta) / \partial G_1({\bf k})$ at $V \to \infty$
if ${\bf k}={\bf 0}$: any of the partial contributions to
$\Sigma({\bf 0},\zeta)$, where zero--vectors
are related to the connecting solid lines in a configuration like
\mbox{\blocks,} is by a
factor $\sim G_1({\bf 0}) \equiv G_{\parallel}({\bf 0})
\simeq M^2 \, V$ larger than the corresponding term of
$\partial \Sigma({\bf 0},\zeta)/\partial G_1({\bf 0})$,
where factor $G_{\parallel}({\bf 0})$ is removed according to derivation of
the connecting lines with ${\bf k}={\bf 0}$. No diverging factor is removed
at $k=+0$ if $G_{\parallel}(+0)$ has a finite value, therefore, $R(+0)$  and
$R({\bf 0})$ are not identical at $n=1$. In any case, this peculiarity refers
only to the longitudinal component, so that
$R_{\perp}(+0) = R_{\perp}({\bf 0})$ holds in general.

The above discussed peculiarities with the extraordinary terms
are irrelevant, as regards the exponents in the asymptotic
expansion of $G_i({\bf k})$. The magnetization $M$ has corrections
to scaling with the same exponents as other terms in~(\ref{eq:viens})
and~(\ref{eq:divi}) (although some expansion coefficients
can be zero). This is consistent with the additional
condition~(\ref{eq:tris}) -- the equation for $M$.
The corrections to scaling have the same
origin at $T<T_c$ and $T>T_c$: the term $ck^2$ in~(\ref{eq:divi})
is by a factor $\sim \Delta^{2 \nu - \gamma}$ smaller than the
leading term and the term ``1'' in~(\ref{eq:wl}) gives
(in the same sense) a correction
$\sim\Delta^{2 \gamma-d \nu}$. In such a way, the general scaling form 
of the solution for $R_i({\bf k})$ and $G_i({\bf k})$ appears to be the same 
at $T<T_c$ and $T>T_c$.
Following the Appendix, it remains unchanged in the complete analysis which
takes into account all possible distributions of ${\bf k}={\bf 0}$ vectors
in the diagrams of~(\ref{eq:ee}), and not only the ``normal'' contributions
without the zero--vector--cumulant insertions.

Thus, for the spational dimensionality $2<d<4$, the correlation function at 
$T<T_c$ has similar singular structure  as in the case of $T>T_c$~\cite{K1}, 
i.~e., we have an asymptotic expansion 
\begin{equation} \label{eq:assi}
G_i({\bf k}) = \sum\limits_{\ell \ge 0} \Delta^{-\gamma+\gamma_{\ell}}
g_i^{(\ell)} \left({\bf k} \Delta^{-\nu} \right)
\end{equation}
with $\gamma_0=0$ and correction exponents
\begin{equation} \label{eq:expo}
\gamma_{\ell}=n_{\ell} \, (2 \nu -\gamma)
+ m_{\ell} \, (2 \gamma - d \nu)  
\end{equation}
valid for $\ell \ge 1$, where $n_{\ell}$ and $m_{\ell}$ are integer numbers
$\ge 0$, and $n_{\ell}+m_{\ell} \ge 1$.
Therefore, using the same arguments as in the case of $T>T_c$~\cite{K1}, 
we conclude that the possible values of exponents $\gamma$ and $\nu$ are given
by~(\ref{eq:resg}) and~(\ref{eq:resn}). It agree with the generalized scaling 
hypothesis in Sec.~\ref{sec:gsh} which tells us that the values of the 
exponents must be the same at $T>T_c$ and $T<T_c$.

\section{The asymptotic long--wave behavior below $T_c$ at $n>1$}
\label{sec:crbn}

\subsection{Discussion of the existing results}
\label{sec:cran}

According to the conventional
believ~\cite{SchwMi,FBJ,BW,WZ,Nel,BZ,Sch,Law1,Law2,Tu,HL},
the transverse Greens function $G_{\perp}({\bf k})$
diverges like $k^{-\lambda_{\perp}}$ with $\lambda_{\perp}=2$ at
$k \to 0$ below $T_c$ for the systems
with $O(n \ge 2)$ rotational symmetry. It corresponds to the
$G_{\parallel}({\bf k}) \sim k^{d-4}$ divergence of the
longitudinal Greens function. Besides, the singular structure
of the correlation functions is represented by an expansion
in powers of $k^{4-d}$ and $k^{d-2}$~\cite{Sch,Law1}.
Formally, our results agree with these ones
at $\lambda_{\perp}=2$.
Nevertheless, below we will show
that $\lambda_{\perp}<2$ holds near two dimensions at $n=2$.

As usually accepted in lattice models, here
(in this subsection) we define that all the parameters of
the normalized Hamiltonian $H/T$~(\ref{eq:H1}) are proportional
to the inverse temperature $1/T$. In this case $r_0$ is negative.
We propose the following argument. The assumption
that $G_{\perp}({\bf k}) \simeq a(T) \, k^{-2}$ holds
(with some temperature--dependent amplitude $a(T)$)
in the stable region below the critical point, i.~e., at
$T \le T_c/C$, where $C$ is an arbitrarily large constant,
leads to a conclusion that the critical temperature $T_c$
continuously tends to zero at $d \to 2$ (supposed $d>2$).
Really, at $\lambda_{\perp}=2$ we have
\begin{eqnarray} \label{eq:fufelis}
&&\left< \varphi^2({\bf x}) \right> = \frac{1}{V} \sum\limits_{i,{\bf k}}
G_i({\bf k}) \simeq M^2 \\
&&+ (2 \pi)^{-d} \left[ \int G'_{\parallel}({\bf k})
d{\bf k} + \frac{(n-1) \, S(d) \Lambda^{d-2} \, a(T)}{d-2} \right] \;,
\nonumber
\end{eqnarray}
where $S(d)$ is the area of unit sphere in $d$ dimensions.
Since the amplitude of the transverse fluctuations never can
vanish at a finite temperature, Eq.~(\ref{eq:fufelis}) implies that
the average $\left< \varphi^2({\bf x}) \right>$  diverges at
$T=T_c/C$ when $d \to 2$, if $T_c$ remains finite. Thus, we obtain an
unphysical result unless the critical temperature $T_c$
and, therefore, $a(T_c/C)$ tend to zero at $d \to 2$.

On the other hand, it is a rigorously stated fact~\cite{TKHT,FS} that 
the classical 2D $XY$ model undergoes the Kosterlitz--Thouless 
phase transition at a finite temperature $T_{\mbox{\tiny KT}}$.
It means that a certain structural order without the spontaneous
magnetization exists within the temperature region
$T < T_{\mbox{\tiny KT}}$. There is a
general tendency of disordering with decreasing the spatial
dimensionality $d$, and not vice versa.
Thus, since the structural order exists at $T < T_{\mbox{\tiny KT}}$
and $d=2$, some kind of order necessarily exists also
at $T < T_{\mbox{\tiny KT}}$ and $d>2$. Since the classical $XY$ model
undergoes the disorder $\rightarrow$ long--range order phase
transition at $d>2$, this obviously is the long--range order.
Thus, the critical temperature at
$d=2 + \varepsilon$ is $T_c \ge T_{\mbox{\tiny KT}} \ne 0$ for an
infinitesimal and positive $\varepsilon$, and it drops to zero by a
jump at $d=2 - \varepsilon$, as consistent with the rigorous
consideration in~\cite{TKHT}. 

The classical $XY$ model belongs to the same universality
class as the actual $\varphi^4$ model at $n=2$, which means that
both models become fully equivalent
after a renormalization (a suitable renormalization will be
discussed in Sec.~\ref{sec:rg}). Thus,
$T_c$ does not vanish at $d \to 2$ (for $d>2$) also in the $\varphi^4$
model. In such a way, the assumption
$G_{\perp}({\bf k}) \simeq a(T) \, k^{-2}$ leads to a contradiction.
In the stable region $T<T_c/C$, the Gaussian approximation
$G_{\perp}({\bf k}) \simeq 1/(2ck^2)$
makes sense at finite not too small values of $k$. 
The above contradiction means that the Gaussian approximation
with $\lambda_{\perp}=2$ cannot be extended to $k \to 0$ in vicinity
of $d=2$. The contradiction is removed only if $\lambda_{\perp}<2$
holds at $d \to 2$ in the actual case of $n=2$.

It has been stated in~\cite{Law1,Tu} that the essentially Gaussian
result $\lambda_{\perp}=2$ of the perturbative RG theory should be
exact.
However, some of the obtained ``exact'' results are rather unphysical. 
In particular, we find from Eq.~(3.6) in~\cite{Law2} and from
the formula $\langle \pi^2 \rangle=-6A/u_0$ given in the line
just below that
\begin{equation} \label{eq:mulkibas}
\langle \pi^2({\bf x}) \rangle = (N-1) \int \frac{d^dq}{q^2}
\end{equation}
holds, where $\pi({\bf x})$ is the transverse $(N-1)$--component
field which, in our notation, is composed of $n-1$ components
labeled by index $j \ne 1$.
Eq.~(\ref{eq:mulkibas}) represents a senseless result,
since $\langle \pi^2({\bf x}) \rangle$
given by this equation diverges at $d \to 2$. It is clear that
$\langle \pi^2({\bf x}) \rangle$ cannot diverge in reality,
as it follows from the Hamiltonian density~(2.1)
in~\cite{Law2} (Hamiltonian~(\ref{eq:H}) in our paper): any field
configuration with diverging $\pi^2({\bf x})$
provides a divergent $\pi({\bf x})$--dependent term
\begin{equation} \label{eq:uh}
\sim  \frac{1}{2} \mid \nabla \pi({\bf x}) \mid^2
+ \frac{u_0}{4!} (\pi^2({\bf x}))^2
\end{equation}
in the Hamiltonian density and,
therefore, gives no essential contribution to the statistical averages.
The result~(\ref{eq:mulkibas}) corresponds to a poor approximation
where the second term in~(\ref{eq:uh}) is neglected.
In a surprising way, based on Ward identities,
authors of~\cite{Law2} and related papers have lost all the purely
transverse diagrams, generated by the term
$\frac{u_0}{4!} (\pi^2({\bf x}))^2$, and stated that this is the
exact result at $r_0 \to - \infty$ as well as at $k \to 0$.
According
to~\cite{Law1,Law2}, the actual transverse term appears to be
hidden in a shifted longitudinal field $\bar s$, which 
is considered as an independent Gaussian variable
(cf.~Eqs.~(3.5) and~(3.6) in~\cite{Law1}). 
Obviously, this is the fatal trivial error which leads to the above
discussed unphysical result~(\ref{eq:mulkibas}), since 
the determinant
of the transformation Jacobian (from $\pi$, $s$ to $\pi$, $\bar s$)
is omitted in the relevant functional integrals.
In this manner the $\varphi^4$ model can be immediately reduced
to the $\varphi^2$ model by considering $s=\varphi^2$ as a new variable! 
Including the Jacobian of the \textit{nonlinear} transformation,
the resulting model, however, remains non--Gaussian.
Since~(\ref{eq:mulkibas}) comes from
\begin{equation}
\langle \pi^2({\bf x}) \rangle = (N-1) \, (2 \pi)^{-d}
\int G_{\perp}({\bf k}) \, d^d k \;,
\end{equation}
the unphysical divergence of $\langle \pi^2({\bf x}) \rangle$
means that the predicted Gaussian form of the transverse correlation
function $G_{\perp}({\bf k})$ is incorrect.
Another aspect is that the method used
in~\cite{Law1,Law2} gives $\lambda_{\perp} \equiv 2$ also at $n=2$
in contradiction with our previous discussion concerning the known
rigorous results for the $XY$ model.

Our consideration does not contradict the conventional statement
(see~\cite{Tu} and references therein)
that the Gaussian spin--wave theory~\cite{Dyson} becomes exact at
low temperatures,
but only in the sense that it holds for any given nonzero
${\bf k}$ at $T \to 0$,
and in the limit $\lim\limits_{k \to 0} \lim\limits_{T \to 0}$
in particular. However, the actual limit of interest is $k \to 0$
or, equally, $\lim\limits_{T \to 0} \lim\limits_{k \to 0}$.
Therefore, contrary to the assertions in~\cite{SchwMi,Tu},
it is impossible to make any rigorous conclusion  
regarding the exponent $\lambda_{\perp}$ (or any related exponent)
based on the fact that the Gaussian spin--wave theory becomes
exact at $T \to 0$.
One has to prove that the limits can be exchanged!
There is a reason to believe that the limits cannot be
exchanged first of all because the critized here treatments with 
exchanged limits lead to \textit{unresolvable} problems at $d \to 2$.
There is also a well studied example -- the $XXX$ quantum 
spin chain, where it is straightforward to see
that the $distance \to \infty$ and $T \to 0$ limits cannot be exchanged
in the correlation function considered there~\cite{BKNS}.

This problem persists in the classical treatment of
the many--particle systems~\cite{Wagner} which, in essence,
is based on the Gaussian spin wave theory at $T \to 0$,
as well as in the hydrodynamical description in~\cite{SchwMi},
where the known results of the Gaussian spin wave theory
[Eqs.~(5.1a) and~(5.1b)] have been implemented 
for a complete description.
The treatment of~\cite{Wagner}, evidently, is not
exact, since it breaks down at $d \to 2$ for the
two--component ($n=2$) vector model (where $T_c$ remains finite 
and we fix the temperature $0<T<T_c$) just like we have
discussed already -- the average $\langle \pi^2({\bf x}) \rangle$
is given by the integral~(6.8) in~\cite{Wagner} which
diverges in this case (supposed
$(2 \pi)^{-3} d^3 k \to (2 \pi)^{-d} d^d k$).

A slightly different perturbative RG approach has been developed
in~\cite{BZ} to analyze the nonlinear $\sigma$ model. In this case
the modulus of $\varphi({\bf x})$ is fixed which automatically
removes the divergence of $\langle \pi^2({\bf x}) \rangle$.
A finite external field $h$ has been introduced there
to make an expansion. The correlation functions
have the power--like singularities of interest only at $h=+0$,
which means that in this case we have to consider the limit
$\lim\limits_{k \to 0} \lim\limits_{h \to 0}$, i.~e., the limit
$h \to 0$ must be taken first at a fixed nonzero $k$ ($p$ in formulae
used in~\cite{BZ}). The results in~\cite{BZ} are not rigorous since 
the expansions used there are purely formal, i.~e., 
they break down in this limit.
Besides, contrary to the approximations in~\cite{BZ}, it should be clear
that the exact
renormalization is a rather nontrivial problem which cannot be reduced
to a finding of only two renormalization constants.
If, e.~g., we make a real--space renormalization
of the Heisenberg model, say, with the scaling factor $s=2$,
then the statistically averaged block--spins of the Kadanoff transformation
(composed of $s^d$ original spins)
do not have a fixed modulus -- simply the original model does not
include the constraint $\mid \varphi({\bf x}) \mid = const$
for the block averages.  It means that the transformation with any
finite $s$ yields a Hamiltonian of form different from the
original one, i.~e., the original Hamiltonian with merely renormalized
coupling constant can never be the fixed--point Hamiltonian.

Another approach, which is based on effective Lagrangians, has been
developed in~\cite{HL}. However, due to several rough (fatal) errors
in Sec.~3 of this paper we cannot appreciate the basic results.
For example, we have found that the transformation
\begin{equation} \label{eq:transf}
v_{\mu}^{\Omega} = \Omega v_{\mu} \Omega^{-1} \;,
\hspace{4ex}
\vec{H}^{\Omega} = \Omega \vec{H}
\end{equation}
brings the Lagrangian
\begin{equation}
\mathcal{L}=\frac{1}{2} D_{\mu} \vec{\phi} D_{\mu} \vec{\phi} + \frac{1}{2}
m^2 \vec{\phi} \vec{\phi} + \frac{1}{4} \lambda (\vec{\phi} \vec{\phi})^2
- \vec{H} \vec{\phi}
\end{equation}
with
\begin{equation}
D_{\mu} \vec{\phi}({\bf x}) = \partial_{\mu} \vec{\phi}({\bf x})
+ v_{\mu}({\bf x}) \vec{\phi}({\bf x})
\end{equation}
back to its original form after a space--independent rotation $\Omega$.
Namely,
\begin{equation}
\mathcal{L} \left( v_{\mu}^{\Omega},\vec{H}^{\Omega},\Omega \vec{\phi}
\right) = \mathcal{L} \left( v_{\mu},\vec{H},\vec{\phi} \right)
\end{equation}
holds, as it can be verified by a simple substitution
taking into account that
$
\Omega \vec{A} \, \Omega \vec{B} = \vec{A} \vec{B}
$
holds for any vectors $\vec{A}$ and $\vec{B}$, since
the rotation of the coordinate system does not change the
scalar product, as well as that
$
\partial_{\mu} \Omega \vec{\phi} = \Omega \, \partial_{\mu} \vec{\phi}
$
obviously is true for a space--independent matrix $\Omega$.
As a consequence, the partition functions obey the equation
\begin{equation} 
Z \left( v_{\mu}^{\Omega},\vec{H}^{\Omega} \right)
=  Z \left( v_{\mu},\vec{H} \right)
\end{equation}
with $v_{\mu}^{\Omega}$ and $\vec{H}^{\Omega}$ defined
in~(\ref{eq:transf}). But Eq.~(3.7) in~\cite{HL} 
disagrees with~(\ref{eq:transf}):
it contains an odd term $\Omega \partial_{\mu} \Omega^{-1}$ 
which reduces to $\partial_{\mu}$ in the case of space--independent
$\Omega$.
The following relevant equations on page 247 contain similar errors,
i.~e., it is easy to verify that they do not hold in the simplest
case of a space--independent rotation.
Some numerical support of this theory can be found in
literature~\cite{DHNN,TY},
where a finite--size scaling within a transient region of
very small fields comparable with $L^{-3}$ 
(where $L$ is the linear size of 3D lattice) have
been considered. Our theoretical predictions, however,
refer to the limit $\lim_{h \to 0} \lim_{L \to \infty}$, therefore the
tests made in~\cite{DHNN,TY} are of little interest here.

Concluding this subsection, it is worthy to mention that the experimental
measurements of susceptibility $\chi$ depending on field $h$ in
isotropous ferromagnets like high--purity polycrystalline 
Ni~\cite{SRS} are incompatible with the conventional RG prediction
$\chi \sim h^{(d-4)/2}$. Moreover, our Monte Carlo results
for 3D $XY$ model discussed in Sec.~\ref{sec:test} are incompatible with 
this prediction, as well.
Thus, according to the given theoretical arguments,
the conventional claims that the Gaussian approximation
is asymptotically exact at $k \to 0$ simply cannot be correct,
and there exist also numerical evidences for this.

\subsection{The leading asymptotic behavior} \label{sec:leading}

Let us now discuss the solution below $T_c$ at small wave vectors $k \to 0$
within our diagrammatic approach.
By analyzing several possibilities we have arrived to a
conclusion that the true physical asymptotic solution for $2<d<4$
and $n>1$ is
\begin{equation} \label{eq:GG}
G_{\perp}({\bf k}) \simeq a \, k^{-\lambda_{\perp}} \;, \hspace{5ex}
G_{\parallel}({\bf k}) \simeq b \, k^{-\lambda_{\parallel}} 
\end{equation}
whith some amplitudes $a$ and $b$, and exponents
\begin{equation} \label{eq:exex}
d/2<\lambda_{\perp}<2  
\hspace{3ex} \mbox{and} \hspace{3ex}
\lambda_{\parallel}=2 \lambda_{\perp}-d \;.
\end{equation}
Only in this case the exponents on the left hand side
of equation~(\ref{eq:Dyson}) for $1/G_i({\bf k})$ coincide with those on
the right hand side, if calculated by the method developed in~\cite{K1}
for the case $T=T_c$. Besides, only at $d/2 < \lambda_{\perp} <2$
we arrive to a solution which coincides both with the scaling
hypothesis~(\ref{eq:scm}) and general renormalization group arguments
discussed further in Sec.~\ref{sec:rg}.

Below we show that~(\ref{eq:GG}) and~(\ref{eq:exex}) really represent
a selfconsistent solution of our equations. According to~(\ref{eq:exex}),
$\lambda_{\parallel}>0$ holds, so that $G_{\parallel}({\bf k})$ diverges at
$k \to 0$. Thus, we have $1/G_{\parallel}(+0)=0$. Eq.~(\ref{eq:tris}) then
implies that $R_{\parallel}(+0)=R_{\parallel}({\bf 0})$. The latter relation
always is true for the transverse components, as already discussed
in Sec.~\ref{sec:Asbel}. Our analysis is based on Eq.~(\ref{eq:divi}),
which at these conditions reduces to
\begin{equation} \label{eq:cetri}
1/ \left( 2G_i({\bf k}) \right) =
ck^2 +R_i({\bf k})-R_i({\bf 0}) \;.
\end{equation}
First let us consider only the ``normal'' contributions without the
zero--vector--cumulant insertions, as explained in Sec.~\ref{sec:low1}
and Appendix. In this case (at the condition~(\ref{eq:exex})), 
the main terms in the asymptotic
expansion of $\Sigma({\bf q},\zeta)$ (at ${\bf q} \ne {\bf 0}$
for any given $\Lambda/q$ considered as independent variable)
are represented by partial
contributions, coming from all diagrams in~(\ref{eq:ee}), where
either all $N_j$ solid lines of the $j$--th diagram
are associated with the transverse
components $G_{\perp}({\bf k})$, or $m$ of $2m \le N_j$ solid lines
which are associated with $G_{\parallel}({\bf k})$ have zero
wave vector.
It is true at $2<d<4$ and~(\ref{eq:exex}), since other terms
provide only small corrections, as discussed further in
Sec.~\ref{sec:correc}.
Since $1/\Sigma({\bf 0},\zeta)=0$
holds at $V = \infty$, only those configurations give a nonvanishing
contribution where nonzero wave vectors are related to the waved lines.
Therefore also maximally one half of all lines associated with the
longitudinal component $i=1$ can have zero wave vector, as regards the 
nonvanishing (at $V \to \infty$) terms related to $\Sigma({\bf q},\zeta)$
with ${\bf q} \ne {\bf 0}$.
This holds because solid lines make closed loops and maximally
each second line of a loop can have zero wave vector, provided that all 
waved and dashed lines, connected to this loop, have nonzero wave vectors.

It is suitable to represent the amplitude $b$ in~(\ref{eq:GG}) as
$b=b' \cdot a^2/M^2$. Then, by normalizing all wave vectors to
the current value of $q$, we find that 
the selfconsistent solution of~(\ref{eq:one}) has the scaled form
\begin{equation} \label{eq:si}
\Sigma({\bf q},\zeta)=a^2 q^{d-2\lambda_{\perp}}
\psi \left( \Lambda/q, \zeta, \lambda_{\perp},b',u,d,n \right)
\end{equation}
which is proven by the method described in detail in~\cite{K1},
taking into account all diagrams of~(\ref{eq:ee}) and only the
main terms of the asymptotic expansion at $k \to 0$. These are
the partial contributions discussed in the paragraph above.
It is supposed also that term "1" in~(\ref{eq:wl}), providing
a small corection at $\lambda_{\parallel}>0$, is neglected.
In the same manner, by normalizing all wave vectors to
the current value of $k$, we arrive to the scaled form
\begin{eqnarray}
\partial \Sigma({\bf q},\zeta)/\partial G_j({\bf k})
&=& V^{-1} a k^{-\lambda_{\perp}} \, Y_{\perp}
\left( {\bf q}/k, \Lambda/k, \zeta, \lambda_{\perp},b',u,d,n \right)
\hspace{2ex} : \hspace{2ex} j \ne 1 \label{eq:s1} \\
\partial \Sigma({\bf q},\zeta)/\partial G_1({\bf k})
&=& V^{-1} M^2 k^{-d} \, Y_{\parallel}
\left( {\bf q}/k, \Lambda/k, \zeta, \lambda_{\perp},b',u,d,n \right)
\nonumber \\ &+&  \delta_{{\bf q},{\bf k}} M^2 \,
\hat Y_{\parallel}
\left( {\bf q}/k, \Lambda/k, \zeta, \lambda_{\perp},b',u,d,n \right) \;,
\label{eq:s2}
\end{eqnarray}
where the term with Kronecker's symbol appears due to the extraordinary
contributions discussed in Sec.~\ref{sec:Asbel}. By virtue of~(\ref{eq:s1}),
(\ref{eq:s2}), and~(\ref{eq:Dast}), (\ref{eq:R}) we obtain also
\begin{eqnarray}
R_{\perp}({\bf k}) &=& a^{-1} k^{\lambda_{\perp}} \,
\bar \phi_{\perp}
\left( \Lambda/k, \lambda_{\perp},b',u,d,n \right) \label{eq:uu} \\
R_{\parallel}({\bf k}) &=& b^{-1} k^{\lambda_{\parallel}} \,
\bar \phi_{\parallel}
\left( \Lambda/k, \lambda_{\perp},b',u,d,n \right) \;. \label{eq:uuu}
\end{eqnarray}
The fact that $R_{\perp}({\bf 0})$ is constant means that there exists
the limit $\lim_{k \to 0} R_{\perp}({\bf k}) =R_{\perp}({\bf 0})$, which,
according to~(\ref{eq:uu}), implies that
$a \Lambda^{-\lambda_{\perp}} R_{\perp}({\bf 0})$ does not depend on
$a$ and $\Lambda$ and thus the scaling function $\bar \phi$ can be
represented as
\begin{equation} \label{eq:phiper}
\bar \phi_{\perp} \left( \Lambda/k, \lambda_{\perp},b',u,d,n \right)
=a \Lambda^{-\lambda_{\perp}} R_{\perp}({\bf 0}) \cdot
(\Lambda/k)^{\lambda_{\perp}} + 1/ \left[ 2 \phi_{\perp}
\left( \Lambda/k, \lambda_{\perp},b',u,d,n \right) \right]
\end{equation}
with another scaling function $\phi_{\perp}$ instead of $\bar \phi_{\perp}$.
Analogous equation for $\bar \phi_{\parallel}$ reads
\begin{equation} \label{eq:phipar}
\bar \phi_{\parallel} \left( \Lambda/k, \lambda_{\perp},b',u,d,n \right)
=b \Lambda^{-\lambda_{\parallel}} R_{\parallel}({\bf 0}) \cdot
(\Lambda/k)^{\lambda_{\parallel}} + 1/ \left[ 2 \phi_{\parallel}
\left( \Lambda/k, \lambda_{\perp},b',u,d,n \right) \right] \;.
\end{equation}
Note that we always consider $\lambda_{\perp}$, but not
$\lambda_{\parallel}$, as an independent argument, therefore some
asymmetry appears in formulae. By substituting Eqs.~(\ref{eq:uu})
to~(\ref{eq:phipar}) and~(\ref{eq:GG}) into~(\ref{eq:cetri}),
and neglecting the correction term $ck^2$, we obtain
\begin{eqnarray}
G_{\perp}({\bf k}) &=& a \, k^{-\lambda_{\perp}} = a \, \phi_{\perp}
\left( \Lambda/k, \lambda_{\perp},b',u,d,n \right) \, k^{-\lambda_{\perp}}
\label{eq:per} \\
G_{\parallel}({\bf k}) &=& b \, k^{-\lambda_{\parallel}} =
b \, \phi_{\parallel} \left( \Lambda/k, \lambda_{\perp},b',u,d,n \right)
\, k^{-\lambda_{\parallel}} \;.
\label{eq:par}
\end{eqnarray}

Up to now we have neglected the zero--vector--cumulant
insertions in the derivation of Eqs.~(\ref{eq:per}) and~(\ref{eq:par}).
However, following the method in Appendix, 
these insertions can only renormalize the scaling functions
$\phi_{\perp}$ and $\phi_{\parallel}$, so that 
Eqs.~(\ref{eq:per}) and~(\ref{eq:par}) hold.

 As discussed in Sec.~\ref{sec:exact}, the
limit $u \to 0$ with simultaneous tending of $k$ to zero like
$k \sim u^r \, k_{crit}(u)$ (where $r>0$) has to be considered
to ensure correct critical exponents. The existence of the
solution for~(\ref{eq:per}) and~(\ref{eq:par}) implies the
existence of the corresponding limits for $\phi_{\perp}$
and $\phi_{\parallel}$. 
Note that these functions do not contain
$u$--dependent factors in our scaling analysis
(which is true also for $\phi$ in Eq.~(42) of~\cite{K1}),
and only weak $u$--dependence (at $u \to 0$) 
can be induced by the integration limits in~(\ref{eq:one}).
Thus, Eqs.~(\ref{eq:per}) and~(\ref{eq:par}) yield
\begin{eqnarray}
\lim\limits_{u \to 0} \phi_{\perp} \left( \Lambda u^{-r} k_{crit}^{-1}(u),
\lambda_{\perp},b',u,d,n \right) &=& B_{\perp}
\left( \lambda_{\perp},b',d,n \right) = 1 \label{eq:Bper} \\
\lim\limits_{u \to 0} \phi_{\parallel} \left( \Lambda u^{-r} k_{crit}^{-1}(u),
\lambda_{\perp},b',u,d,n \right) &=& B_{\parallel}
\left( \lambda_{\perp},b',d,n \right) = 1  \;. \label{eq:Bpar}
\end{eqnarray}
Eqs.~(\ref{eq:Bper}) and~(\ref{eq:Bpar}) can be, in principle,
solved with respect to $\lambda_{\perp}$ and $b'$. It follows
herefrom that not only the exponent $\lambda_{\perp}$, but also
the quantity $b'=bM^2/a^2$ is universal, i.~e., dependent only on the
spatial dimensionality $d$ and dimensionality of the order
parameter $n$. In general, no universality of amplitudes is expected,
so that the latter rather surprising conclusion
refers only to the actual limit $u \to 0$. Nevertheless, the universality
of $bM^2/a^2$ coincides with some general non--perturbative
renormalization group arguments discussed in Sec.~\ref{sec:rg}, which
show that the restriction to small $u$ values is purely formal.

\subsection{Corrections to scaling}  \label{sec:correc}

In Sec.~\ref{sec:leading} we have considered only the dominant terms
in the asymptotic solution at $k \to 0$. Now we shall discuss
corrections to scaling. There are following sources of corrections.
\begin{itemize}
\item[(i)]
Since $R_{\perp}({\bf k})-R_{\perp}({\bf 0}) \propto a^{-1}
k^{\lambda_{\perp}}$ and $\lambda_{\perp}<2$ hold, the term
$ck^2$ in Eq.~(\ref{eq:cetri}) for the transverse components $i \ne 1$
produces a correction which is by factor\linebreak
$\varepsilon_1(k) \propto
a \, k^{2-\lambda_{\perp}}$ smaller than the main term at $k \to 0$.
In analogy, a correction $\varepsilon'_1(k) \propto b \,
k^{2-\lambda_{\parallel}}$ is generated in the same equation for $i=1$.
\item[(ii)]
According to~(\ref{eq:si}) and~(\ref{eq:exex}), at any given $\Lambda/q$,
the term "1" in Eq.~(\ref{eq:wl}) represents a small correction 
$\propto a^{-2} q^{\lambda_{\parallel}}$ to the amplitude of the main term.
Finally, it generates an amplitude correction 
$\varepsilon_2(k) \propto a^{-2} k^{\lambda_{\parallel}}$
in the asymptotic expansion of $G({\bf k})$.
\item[(iii)]
Consider partial contributions to
$\Sigma({\bf q},\zeta)$, coming from all diagrams in~(\ref{eq:ee}),
where less than one half of the solid lines which belong to loops
with $i=1$ have zero wave vector. As compared to the dominant contributions
discussed in Sec.~\ref{sec:leading}, they
generate small corrections represented by an expansion in terms of
$\varepsilon_3(k)$, where $\varepsilon_3(k) \propto (b/a) \,
k^{\lambda_{\perp}-\lambda_{\parallel}}$ corresponds to a replacement
of $G_{\perp}({\bf k})$ with $G_{\parallel}({\bf k})$ for one solid
line with nonzero wave vector.
\end{itemize}
Note that the corrections $\varepsilon_{\ell}(k)$ are small at
$k \to 0$ only for $d/2 < \lambda_{\perp} <2$, so that our
analysis cannot be formally extended outside of this interval.
We have included an explicit dependence of $\varepsilon_{\ell}(k)$ on
the amplitudes $a$ and $b$, since their singular behavior 
is relevant for joining of the asymptotic solutions at $T \to T_c$.
Since $\varepsilon'_1(k) \propto \varepsilon_1(k) \, \varepsilon_3(k)$
holds, we have no more than three independent correction sources.

The expansion in powers of $\varepsilon_{\ell}(k)$ is acompanied
by scaling functions depending on $\Lambda/k$.
Like in~(\ref{eq:Bper}) and (\ref{eq:Bpar}), these scaling
functions can be replaced by amplitudes which are independent of $k$,
when considering the limit $u \to 0$ and $k \sim u^r \, k_{crit}(u)$.
This replacement is analogous to that at $T=T_c$ and has
the same motivation~\cite{K1}.
It results in the asymptotic expansion for the Greens function
\begin{equation} \label{eq:asex}
G_i({\bf k}) = \sum\limits_{\ell \ge 0} b_i(\ell) \, k^{-\lambda_i(\ell)} \;,
\end{equation}
where relations $\lambda_i(0) \equiv \lambda_{\parallel}$, $b_i(0) \equiv b$
hold for $i=1$, and 
$\lambda_i(0) \equiv \lambda_{\perp}$, $b_i(0) \equiv a$ -- for
$i \ne 1$. A term with $\ell \ge 1$ represents a correction of order
$\varepsilon_1^{n_1(\ell)} \varepsilon_2^{n_2(\ell)}
\varepsilon_3^{n_3(\ell)}$
with the exponent
\begin{equation} \label{eq:asexpo}
\lambda_i(\ell)= \lambda_i(0)-n_1(\ell) \cdot (2-\lambda_{\perp})
-n_2(\ell) \cdot \lambda_{\parallel}
-n_3(\ell) \cdot (\lambda_{\perp}-\lambda_{\parallel}) \;,
\end{equation}
where $n_j(\ell) \ge 0$ are integer numbers such that
$\sum_j n_j(\ell) \ge 1$. Note that we always allow a possibility
that some of the expansion coefficients, in this case
some of $b_i(\ell)$, are zero. The expansion in powers of
$\varepsilon_2(k) \propto k^{4-d}$ and
$\varepsilon_3(k) \propto k^{d-2}$, proposed by the perturbative RG
theory~\cite{Sch,Law1}, is recovered by formally setting
$\lambda_{\perp}=2$.

\subsection{Joining of asymptotic solutions} \label{sec:join}

Consider now how our expansion~(\ref{eq:asex}) coincides
with~(\ref{eq:scm}) and~(\ref{eq:assi}) when approaching the
critical point. Retaining only the leading terms,
the consistency is ensured
if the scaling functions of the dominant terms behave as
$g_{\perp}(z) \propto z^{-\lambda_{\perp}}$ (for $i \ne 1$) and
$g_{\parallel}(z) \propto z^{-\lambda_{\parallel}}$ (for $i=1$)
at $z \to 0$, and
$a \propto \hat \xi^{\lambda-\lambda_{\perp}}$,
$b \propto \hat \xi^{\lambda-\lambda_{\parallel}}$
hold for the amplitudes in~(\ref{eq:GG})
at $\Delta \to 0$, where $\lambda=\gamma/\nu=2-\eta$. Here
$\hat \xi= \Delta^{-\nu}$ is an analog of the correlation length.
This is a property of the solution exceptionally at
$d/2 < \lambda_{\perp} <2$ that the consideration of the long--wave
limit does not provide any constraint for the amplitude $a$ in~(\ref{eq:per}),
whereas the other amplitude $b$ is related to $a$ via $b=b' \cdot a^2/M^2$.
Taking into account the scaling law~(\ref{eq:scbeta}), the relations
$a \propto \hat \xi^{\lambda-\lambda_{\perp}}$ and
$b \propto \hat \xi^{\lambda-\lambda_{\parallel}}$
mean that $bM^2/a^2$ is constant at
$\Delta \to 0$. It is consistent with the statement in Sec.~\ref{sec:leading}
that $b'=const$ at $u \to 0$ for any $T<T_c$.

The expansion~(\ref{eq:assi}) with the exponents
$\gamma_{\ell}$~(\ref{eq:expo})
completely agree with~(\ref{eq:asex}) and~(\ref{eq:asexpo}) provided
that the scaling functions have an asymptotic expansion 
\begin{equation}
g_i^{(\ell)}(z) = \sum\limits_{\ell \ge 0}
B_i^{(\ell)}  z^{-\lambda_i(\ell)}
\end{equation}
at $z \to 0$. In this case the amplitudes
$b_i(\ell) \sim \Delta^{-\gamma+\gamma_{\ell}+\nu \lambda_i(\ell)}$
have corrections to scaling where the main term is multiplied by
$\propto \Delta^{\gamma_m}$.

\subsection{Non--perturbative renormalization group arguments}
\label{sec:rg}

The relation $\lambda_{\parallel}=2 \lambda_{\perp}-d$ as well as the
universality of the ratio $bM^2/a^2$ have a simple interpretation
in view of some renormalization group (RG) analysis. 
Our $\varphi^4$ model can be formulated on a discrete lattice,
representing the gradient term by finite differences.
At $h = +0$, we consider the transformation consisting of
\begin{itemize}
\item[(i)] Kadanoff transformation replacing single lattice spins
by block--spins, where each block--spin is an average over $s^d$ spins;
\item[(ii)] shrinkage of  the new lattice $s$ times to return
to the initial lattice constant. In distinction to
the standard renormalization, we do not rescale the 
field $\varphi$. 
\end{itemize}
The distribution over block--spins is described by
new Hamiltonian $T_s H$, where the notation $T_s$ is used
to distinguish from the standard RG transformation
$R_s$. The Kadanoff transformation does not change neither
the magnetization nor the long--distance behavior of the
real--space Greens functions
$\widetilde G_{\perp}({\bf x}) =
\left< \varphi_i({\bf x}_1) \varphi_i({\bf x}_1+{\bf x}) \right>=
\hat a x^{\lambda_{\perp}-d}$ ($i \ne 1$)
and $\widetilde G_{\parallel}({\bf x}) =
\left< \varphi_1({\bf x}_1) \varphi_1({\bf x}_1+{\bf x}) \right>-M^2 =
\hat b x^{\lambda_{\parallel}-d}$  at $x \to \infty$,
where $\hat a = c_a \cdot a$ and $\hat b = c_b \cdot b$ are the amplitudes.
The proportionality coefficients $c_a$ and $c_b$
are universal, since $G_{\perp}({\bf k}) \simeq a \, k^{-\lambda_{\perp}}$
and $G_{\parallel}({\bf k}) \simeq b \, k^{-\lambda_{\parallel}}$
are the Fourier transforms of $\widetilde G_{\perp}({\bf x})$
and $\widetilde G_{\parallel}({\bf x})$, respectively.
Taking into account the shrinkage of the lattice at step (ii),
magnetization $M$ is invariant of the transformation $T_s$, whereas the
amplitudes rescale as
$\hat a(s) = \hat a(1) \cdot s^{\lambda_{\perp}-d}$ and
$\hat b(s) = \hat b(1) \cdot s^{\lambda_{\parallel}-d}$.
The modulus conservation principle is true at large renormalization
scales $s$, since the variation of average modulus for large blocks of
spins is related to a much greater increase in the systems energy
as compared to a gradual long--wave perturbation of spin orientation.
The validity of this principle is restricted by the condition
that the mean amplitude of the relative fluctuation of the modulus 
has to be much smaller than the mean squared
fluctuation of the orientation angle for the block--spins of 
the Kadanoff transformation.
The non--perturbative renormalization group arguments given below
are in agreement with our foregoing diagrammatic analysis, 
assuming that this condition is fulfilled for large $s$ 
in the actual case of $2<d<4$.
Thus, the renormalized Hamiltonian can be written as
\begin{equation} \label{eq:Ts}
T_s \, (H/T) \simeq \sum\limits_{\bf x} \left[
\frac{1}{\mid \delta (\mid \varphi({\bf x}) \mid - \varphi_0 ) \mid}
+ Q \left\{ \hat a^{-1/2}(s) \, \varphi_{\perp}({\bf x}) \right\} \right]
\end{equation}
at $s \to \infty$, where the first term represents the modulus
conservation principle allowing only those configurations
with non--diverging Hamiltonian where
$\mid \varphi({\bf x}) \mid =\varphi_0$,
and $Q$ is some functional of the configuration of the transverse order
parameter ($n-1$ component vector) field $\varphi_{\perp}({\bf x})$.
In this case only the infinitely small (at $s \to \infty$)
transverse components $\varphi_i({\bf x})$ with $i \ne 1$
are independent variables, since 
\begin{equation} \label{eq:modc}
\varphi_0 - \varphi_1({\bf x})  \simeq \sum\limits_{i=2}^n
\varphi_i^2 ({\bf x}) /(2M) 
\end{equation}
holds according to $\mid \varphi({\bf x}) \mid = \varphi_0$.
Not loosing the generality, we have considered the spatial configuration
of the normalized transverse components
$\hat a^{-1/2}(s) \, \varphi_{\perp}({\bf x})$
as an argument of the functional $Q$. 
According to the definition of $\widetilde G_{\perp}({\bf x})$,
the function
$f_{\perp}({\bf x})= \left<
\hat a^{-1/2}(s) \, \varphi_i({\bf x}_1) \cdot
\hat a^{-1/2}(s) \, \varphi_i({\bf x}_1+{\bf x}) \right>
= \hat a^{-1}(s) \widetilde G_{\perp}({\bf x})$
with $i \ne 1$ has a universal asymptotic behavior
$f_{\perp}({\bf x})=x^{\lambda_{\perp}-d}$ at $x \to \infty$.
Since this average is composed of arguments of $Q$,
the sufficient condition for its universal asymptotic
behavior is the universality of the functional $Q \{ z({\bf x}) \}$.
The latter is consistent with the idea that the transformation
of $Q$ (assuming that at any $s$ the transformed Hamiltonian
can be approximated by~(\ref{eq:Ts}) according to some a priori
defined criterion) has a fixed point
\begin{equation} \label{eq:fixp}
Q^* \{ z({\bf x}) \} = \lim\limits_{s \to \infty} T_s Q \{ z({\bf x}) \} 
\end{equation}
which, however, may be different for each universality class.
In the conventional RG transformation $R_s$ the field would be rescaled
as $\varphi_{\perp}({\bf x}) s^{(d-\lambda_{\perp})/2} \Rightarrow 
\varphi_{\perp}({\bf x})$, so that $Q$ in~(\ref{eq:Ts}) would
contain no explicit scaling factor $s$.
Nevertheless, we prefer our notation,
since it is suited to express the modulus conservation principle.
Accepting~(\ref{eq:fixp}), any statistical average composed of
arguments $\hat a^{-1/2}(s) \, \varphi_{\perp}({\bf x})$ is universal
at $s \to \infty$. In particular,
$f_{\parallel}({\bf x})= \left<
\left( \hat a^{-1/2}(s) \, \varphi_i({\bf x}_1) \right)^2 \cdot
\left( \hat a^{-1/2}(s) \, \varphi_i({\bf x}_1+{\bf x}) \right)^2 \right>$
with $i \ne 1$ is a universal function. According to~(\ref{eq:modc}),
we have
$\widetilde G_{\parallel}({\bf x})= \hat b(s) x^{\lambda_{\parallel}-d}
=(n-1)(2M)^{-2} \hat a^2(s) \,
\left[ f_{\parallel}({\bf x}) - f_{\parallel}({\bf \infty}) \right]$
at $x \to \infty$ and $s \to \infty$.
The universality of $f_{\parallel}({\bf x})$
then implies that $\hat b(s) M^2/\hat a^2(s)$ must be universal at
$s \to \infty$. According to the scaling rules
$\hat a(s) = \hat a(1) \cdot s^{\lambda_{\perp}-d}$ and
$\hat b(s) = \hat b(1) \cdot s^{\lambda_{\parallel}-d}$,
the latter is possible only if $\lambda_{\parallel}=2 \lambda_{\perp}-d$
holds, whence it follows also that $\hat b(1) M^2/\hat a^2(1)$ and
$b(1) M^2/ a^2(1) \equiv b M^2/ a^2$ are universal, i.e., dependent
merely on $n$ and $d$.
Thus we recover one of relations~(\ref{eq:exex}),
as well as the universality of the ratio $b M^2/ a^2$
discussed in Sec.~\ref{sec:leading}.

It is very likely that the accuracy of~(\ref{eq:Ts}) is limited 
even at $s \to \infty$. However, there exists a less constrained form 
\begin{equation} \label{eq:Ts1}
T_s \, (H/T) = \sum\limits_{\bf x} 
 Q \left\{ \left( M-\varphi_1({\bf x}) \right) M \hat a^{-1}(s), \,  
\hat a^{-1/2}(s) \, \varphi_{\perp}({\bf x}) \right\} 
\end{equation}
for the renormalized Hamiltonian at $s \to \infty$, as consistent with the idea 
that Eq.~(\ref{eq:modc}) with
$\varphi_0 = M + {\cal O} \left( \varphi_{\perp}^2/M \right)$
holds approximately for relevant configurations of block--spins. 
It also leads to the relation
$\lambda_{\parallel}=2 \lambda_{\perp} - d$ and the universality of $b M^2/ a^2$.

Contrary to the previous discussion in Sec.~\ref{sec:leading}, in this case
our conclusions are not restricted to small $u$.

\subsection{Magnetization in a small external field} \label{sec:mag}

Here we discuss the qualitative behavior of magnetization $M$
in a small external field.

The modulus conservation principle holds far below the critical point
(at large negative $r_0$) where the fluctuations of modulus
$\mid \varphi({\bf x}) \mid$ are reduced to a small vicinity
of $\varphi_0(h) \simeq \sqrt{-r_0/(2u)}-h/(4r_0)$,
as consistent with the minimum of Hamiltonian~(\ref{eq:H}).
In this case we have
\begin{equation} \label{eq:Mh}
M(h)= \langle \varphi_1({\bf x}) \rangle
\simeq \varphi_0(h) \, \left( 1- \frac{1}{2}
\left< \theta^2({\bf x}) \right>(h) \right)  \;,
\end{equation}
where $\theta({\bf x})$ is the angular deviation of $\varphi({\bf x})$
from the direction of the external field ${\bf h}$.
We consider the limit
$\lim\limits_{c \to \infty} \lim\limits_{r_0 \to -\infty}
\lim\limits_{h \to 0}$
where Eq.~(\ref{eq:Mh}) is asymptotically exact, since
$\mid \varphi({\bf x}) \mid = const$ holds with an unlimited
accuracy and, simultaneously, the angular fluctuations are suppressed.
The variation of $\varphi_0(h)$ with $h$ is analytical, whereas the
singular behavior of $M(h)$ at $h \to 0$ is due to the term
\begin{equation}
\left< \theta^2({\bf x}) \right>(h) \sim
\widetilde G_{\perp}({\bf 0})= (2 \pi)^{-d}
\int G_{\perp}({\bf k}) \, d {\bf k} \;.
\end{equation}
The transverse correlation function behaves like
$G_{\perp}({\bf k}) \simeq a \, k^{-\lambda_{\perp}}$
(the $k \to 0$ asymptotic at $h=0$) when $k$ is decreased below
some $k_{crit}$ (defined for any given $r_0$ and $c$) if $h \to 0$
until it saturates at the known value $M/h$ valid for
${\bf k}={\bf 0}$. From this we find
\begin{equation} \label{eq:mag}
M(h)-M(+0) \propto h^{(d/\lambda_{\perp})-1}
\hspace{4ex} \mbox{at} \hspace{2ex} h \to 0 \;.
\end{equation}
Since the exponent $\rho=(d/\lambda_{\perp})-1$ is universal,
Eq.~(\ref{eq:mag}) is valid
for any $T<T_c$ including vicinity of the critical point.
This yields the longitudinal susceptibility
\begin{equation}
\chi_{\parallel} = \partial M(h)/\partial h \propto
h^{(d-2 \lambda_{\perp})/ \lambda_{\perp}}
= h^{-\lambda_{\parallel} / \lambda_{\perp}}
\hspace{4ex} \mbox{at} \hspace{2ex} h \to 0 \;.
\end{equation}
According to~(\ref{eq:exex}), we have $(d/2)-1 <\rho< 1$, which 
yields $1/2 < \rho <1$ in three dimensions.
The lower value $\rho=0.5$ corresponds to the conventional
statement~\cite{PaPo,FBJ,Law2} that $\lambda_{\perp}=2$.
Our numerical test in the following section, however, 
does not support this possibility.

\section{Monte Carlo test in 3D $XY$ model}
\label{sec:test}

To verify the theoretical predictions for the exponent 
$\rho=(d/\lambda_{\perp})-1$ in~(\ref{eq:mag}), we have made Monte Carlo (MC)
simulations of 3D $XY$ model on simple cubic lattice with the Hamiltonian
\begin{equation}
\frac{H}{T}=-K \left( \sum\limits_{\langle i j \rangle}
{\bf s}_i {\bf s}_j + \sum_i {\bf h s_i} \right) \;,
\end{equation} 
where ${\bf s}_i$ is the spin variable (two--component vector) of
the $i$--th lattice site, $K$ is the 
coupling constant, and ${\bf h}$ is the external field.
Based on the universality argument (cf.~Sec.~\ref{sec:rg}),
the same value of $\rho$ is valid also for the actual
$\varphi^4$ model with two--dimensional order parameter ($n=2$).
The simulations have been made in the ordered phase at 
$K=0.475, 0.5, 0.55 >K_c$, where $K_c \simeq 0.4542$~\cite{SM} is the critical point.
Only the case $K=0.5$ is discussed in detail, since the analysis
made at $K=0.475$ and $K=0.55$ is similar. 
The quatity $\langle \mid m \mid \rangle$, where $m$ is the magnetization
per spin, has been evaluated for different linear sizes of the lattice $L$. 
The Wolff's cluster algorithm~\cite{Wolff} has been used
at $h=0$. The results are $\langle \mid m \mid \rangle =$ 0.570297(23),
0.542411(21), 0.530317(20), 0.524606(19), 0.521846(26), and 0.520449(35)
for $L=$ 8, 16, 32, 64, 128, and 256, respectively.  
These values for $L=8$ to $256$ have been obtained by an averaging over,
respectively, $8 \cdot 10^7$, $2 \cdot 10^7$,
$5 \cdot 10^6$, $1.25 \cdot 10^6$, $3.125 \cdot 10^5$, and $7.5 \cdot 10^4$
cluster algorithm steps. Each simulation has been split typically in 51
bins to calculate the mean value and the standard deviation discarding
the first bin (first 2 bins at $L=256$). The value at $L=256$ has been obtained by a
weighted  averaging
over two simulations including totally 30 ($10 + 20$) not discarded bins, 
each consisting of 2500 cluster updates. 
 
The simulations at $h>0$ have been done by the Metropolis algorithm.
The results of simulation for $L=8$, $16$, $32$ and $64$ are listed in Tab.~\ref{tab}. 
\begin{table}
\caption{\small The MC simulated values of $\langle \mid m \mid \rangle$ for 3D $XY$ model
depending on the external field at a fixed coupling constant $K=0.5$. }
\label{tab}
\begin{center}
\begin{tabular}{|c|l|l|l|l|}
\hline
& \multicolumn{4}{|c|}{\rule[-2.5mm]{0mm}{7mm} $\langle \mid m \mid \rangle$} \\ \cline{2-5}
\raisebox{1.5ex}{h} 
& \rule[-2mm]{0mm}{4mm}
    L=8               & L=16          & L=32         & L=64         \\ \hline
0.028 & 0.586155(85)  & 0.567394(63)  & 0.562201(78) & 0.561218(45) \\ 
0.04  & 0.593381(86)  & 0.576368(91)  & 0.572118(70) & 0.571425(55) \\
0.056 & 0.601719(100) & 0.586635(99)  & 0.583529(77) & 0.583020(45) \\
0.08  & 0.613203(93)  & 0.600377(109) & 0.597872(63) & 0.597552(52) \\
0.112 & 0.626239(89)  & 0.615581(81)  & 0.613741(61) & 0.613527(46) \\
0.16  & 0.643071(78)  & 0.634658(63)  & 0.633304(50) & 0.633242(31) \\
0.224 & 0.661554(58)  & 0.655097(58)  & 0.654166(34) & 0.654016(28) \\
0.32  & 0.684105(62)  & 0.679169(51)  & 0.678533(31) & 0.678483(22) \\
0.448 & 0.707654(44)  & 0.704018(37)  & 0.703606(32) & 0.703496(20) \\
0.64  & 0.734667(38)  & 0.732058(25)  & 0.731754(22) & 0.731714(16) \\ \hline
\end{tabular}
\end{center}
\end{table}
The statistical averages have been evaluated from $3.2 \cdot 10^6 \times (64/L^2)$ sweeps at
$0.08 \le h \le 0.64$, discarding no less than $50 \, 000$ sweeps from the
beginning of the simulation to ensure an accurate equilibration.
The total length of the simulation as well as the discarded part have been increased
by a factor $0.08/h$ at $h<0.08$.
Like in the case of $h=0$, each simulation has been split in bins, using the last 50
ones for the estimations. 

The linear congruatial generator with multiplier $7^5$ and modulo
$2^{31}-1$ (Lewis generator), improved by a shuffling, has been used as a
source of pseudo random numbers. The standard 
shuffling scheme~\cite{MC} with the length of string $N=10^6$ has been completed
by a second shuffling, where the whole cycle (consisting of $2^{31}-2$ numbers)
of the original generator has been split in $2^{20}$ segments,
restarting the generation from the beginning of a new randomly choosen
segment when the previous one is exhausted.
The scheme provided excellent results in test simulations of 2D Ising model,
where the simulated values of internal energy, specific heat $C_V$,
and its first two derivatives have been compared with the exact results.  
No systematic deviations have been observed in rather long
simulations at the critical point providing the standard error in $C_V$ as small 
as $0.02\%$ for $48 \times 48$ lattice and $0.11\%$ for $256 \times 256$ lattice.  

\begin{figure}
\centerline{\psfig{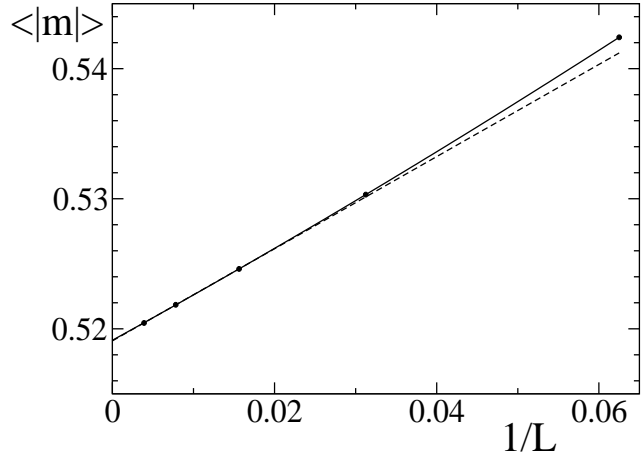}}
\caption{The mean magnetization modulus $\langle | m | \rangle$
vs the inverse linear system size $1/L$ at $K=0.5$, $h=0$. 
The statistical errors are much smaller than the symbol size. 
The linear extrapolation (dashed line) and the quadratic fit 
(solid line) yield the asymptotic values $0.519073(37)$ and $0.519116(30)$,
respectively.}
\label{h00}
\end{figure}

\begin{figure}
\centerline{\psfig{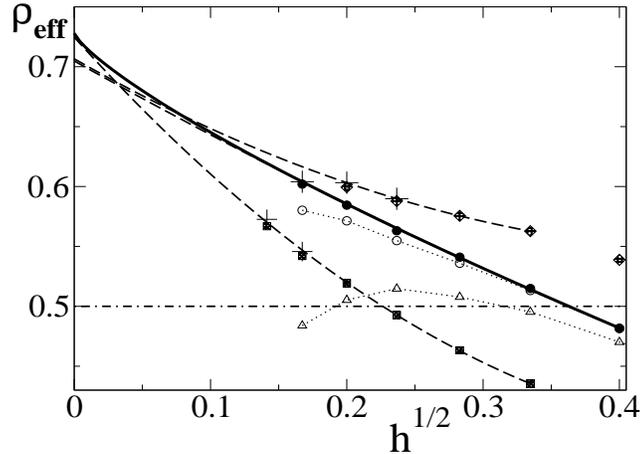}}
\caption{The effective exponent
$\rho_{eff}(h,L)$ for $K=0.5$ evaluated at $L=16$ (triangles), 
$L=32$ (empty circles), and $L=64$ (solid circles). The squares
and rhombs show the $L=64$ results for two other couplings (temperatures)
$K=0.475$ and $K=0.55$, respectively. The estimates for $L= \infty$, 
obtained by correcting the $L=64$ values, are swown by pluses.
The statistical errors are within the
symbol size. The dashed and solid lines show the fits 
$\rho_{eff}(h)= \rho + a_1 h^{1/2}+ a_2 h$ and
$\rho_{eff}(h)= \rho + a h^{\omega}$, respectively.
The horizontal dot--dashed line shows the asymptotic
value of the ``exact'' RG theory.
}
\label{expo}
\end{figure}

The quantity $M(+0)$ in~(\ref{eq:mag}) has been evaluated by extrapolating our
$\langle \mid m \mid \rangle$ data to the thermodynamic limit, based on an
empirical observation that $\langle \mid m \mid \rangle$ is almost linear
function of the inverse lattice size $1/L$, as it is evident from
Fig.~\ref{h00}. 
The linear fit of three largest sizes (dashed line in
Fig.~\ref{h00}) provides $M(+0)=0.519073(37)$, the quadratic fit
within $16 \le L \le 256$ yields $M(+0)=0.519116(30)$ (solid line),
whereas the cubic fit within $8 \le L \le 256$ gives us $M(+0)=0.519096(33)$.
The values of goodness $Q$ (see~\cite{Recipes} for the definition) of the respective  
fits are $0.769$, $0.767$, and $0.794$. 
We have accepted the result $M(+0)=0.519116(30)$
of the quadratic fit for our further estimations, although we could choose
any of two other values as well, since the differences are rather small.
Moreover, even a shift by $20$ standard deviations would not change the qualitative
picture in Fig.~\ref{expo}, where the effective critical exponent $\rho_{eff}(h,L)$
is shown, defined as the mean slope of the $\ln [M(h',L)-M(+0)]$ vs $\ln h'$ plot
within $h' \in [h,2h]$, where $M(h',L)$ is the value of $\langle \mid m \mid \rangle$
at the field $h'$ and the linear lattice size $L$.
Thus, the effective exponent is given by
\begin{equation}
\rho_{eff}(h,L) = \ln \left[ \frac{M(2h,L)-M(+0)}{M(h,L)-M(+0)} \right]
/ \ln 2 \;,
\end{equation}
and the true value of the critical exponent $\rho$
is obtained in the limit $\rho = \lim\limits_{h \to 0} \lim\limits_{L \to \infty}
\rho_{eff}(h,L)$.
As we see from Fig.~\ref{expo}, the effective exponent $\rho_{eff}(h,L)$
increases monotoneously and converges to a certain value
$\rho_{eff}(h) = \lim\limits_{L \to \infty} \rho_{eff}(h,L)$
with increasing of $L$. The data in Tab.~\ref{tab} obey an approximate
scaling relation 
$\rho_{eff}(h,2L)-\rho_{eff}(h,L) =$ \linebreak
$A \left[ \rho_{eff}(4h,L)-\rho_{eff}(4h,L/2) \right]$
which holds with an almost constant value of coefficient $A$,
i.~e., $A \approx 1.15$ for $L=16$ and $A \approx 1.2$ for $L=32$.
According to the finite--size scaling theory $L/\xi$ is the only essential
scaling argument. It implies that the correlation length $\xi$ is roughly
proportional to $h^{-1/2}$ within $h \in [0.028;0.64]$. At $h=0.112$
our results for $L=32$ closely agree with those at $L=64$, which indicates
that $\xi$ is several times smaller than $32$ in this case. 
The actual scaling analysis then implies that $\xi$ is several times
smaller than $L=64$ at $h=0.028$. The scaling relation we found allows us
to evaluate $\rho_{eff}(h,64)-\rho_{eff}(h)$ for $h \ge 0.028$.
It leads to the conclusion that, 
with a high enough accuracy, our results at $L=64$ (solid circles in Fig.~\ref{expo})
correspond already to the thermodynamic limit at not too small fields $h \ge 0.04$,
whereas a small correction (increment) about $0.002$ is necessary  
to get the true value of $\rho_{eff}(h)$ at $h=0.028$. This value is indicated
in Fig.~\ref{expo} by a plus. Even larger than $L=64$ lattices have
to be considered for a reliable estimation of the thermodynamic limit
at $h=0.02$ an smaller fields.

 According to the ``exact'' RG theory critised in Sec.~\ref{sec:cran}, one can
expect that $\rho_{eff}(h)$ converges to the asymptotic value
$\rho = 0.5$ linearly in  $h^{1/2}$ at $h \to 0$,
as consistent with the expansion of $M$ in powers of $h^{1/2}$.
We have used this scale in Fig.~\ref{expo} to show that the $\rho_{eff}(h)$
vs $h^{1/2}$ plot at $K=0.5$ (solid circles at $h>0.28$ and plus at $h=0.28$) 
goes, indeed, almost linearly, but clearly not to the ``exact'' value $0.5$. 
Two different fits we made, namely,
$\rho_{eff}(h)= \rho + a_1 h^{1/2}+ a_2 h$ (dashed curve) 
and $\rho_{eff}(h)= \rho + a h^{\omega}$ (solid curve) yield
$\rho = 0.705(9)$ and $\rho = 0.728(25)$ (with $\omega = 0.394(50)$),
respectively. The fits of the first kind have been made also at $K=0.475$ and $K=0.55$ 
(lower and upper dashed lines) providing $\rho = 0.725(9)$ and $\rho=0.707(36)$, 
respectively. The fits of the second kind are not stable enough in these cases because of
too large inaccuracy in $\omega$. The above listed estimates of $\rho$
satisfactory well agree with the average value about $0.716$ and, thus, confirm
the expected universality of this exponent.
A similar method of effective exponents has been tested in 2D Ising model~\cite{K3} 
where it provided very accurate results.
 
The actual results for $XY$ model agree with our prediction $1/2 < \rho <1$
for 3D case (cf.~Sec.~\ref{sec:mag}) and are incompatible with 
the conventional believe that $\rho$ should be $1/2$.
To the contrary, it has been claimed in~\cite{EM,EHMS} that the MC simulated
magnetization data for $O(4)$ and $O(2)$ models well agree with the predictions 
of the Gaussian spin wave theory.  
However, we failed to see any serious argument in these papers by J. Engels et. al,
since the only quantity which could be extracted from the magnetization
data and precisely compared to the theory, i.~e., the universal exponent $\rho$,
has not been evaluated there. Moreover, their magnetization plots for the $O(2)$ 
model are remarkably nonlinear functions of $h^{1/2}$, i.e.,
they do not provide even an indirect evidence that $\rho$ is just $1/2$.
The only values we found in~\cite{EHMS}, which can be compared to ours, are 
$M(+0)=0.5186(01)$ at $K=0.5$ and $M(+0)=0.6303(1)$ at $K=0.55$.
Our respective values $0.519116(30)$ and $0.630545(24)$ are similar. 
Since there is no reason to assume that one of us has made a wrong simulation, 
the small but remarkable (5 their standard deviations at $K=0.5$) discrepancies, 
obviously, are due to the different extrapolation procedures used.
Our values are more precise and reliable, since our method allows to 
estimate $M(+0)$ from simulations at $h=0$ without any extrapolation
over $h$ used in~\cite{EHMS}. Besides, we have used a larger maximal system 
size $L=256$ as compared to $L=160$ in~\cite{EHMS}.
As a result, the extrapolation gap in $M$
is by an order of magnitude smaller in our case.
The spontaneous magnetization $M(+0)$ has been evaluated in~\cite{EHMS}
from the fit $M(h)=M(+0)+ah^{\rho}+bh$ with $\rho \equiv 1/2$.
In such away, the extrapolation which is biased by $\rho=1/2$ gives
systematically underestimated values of $M(+0)$ thus providing
an indirect evidence that $\rho \ne 1/2$.
If $\rho=1/2$ is replaced with $\rho>1/2$ in this ansatz, then the
extrapolation gap becomes smaller ($M(+0)$ becomes larger) and the above
discussed small discrepancies in $M(+0)$ can be removed. This analysis suggests 
that $\rho>1/2$ holds, as consistent with the direct estimation in our paper.

\section{Conclusions}

In the present work we have extended our diagrammatic method
introduced in~\cite{K1} to study the $\varphi^4$ model in the ordered
phase below the critical point, i.~e., at $T<T_c$.
In summary, we conclude the following.
\begin{enumerate}
\item  The diagrammatic equations derived in~\cite{K1} have been
generalized to include the symmetry breaking term fixing the axis
of ordering at $T<T_c$~(Sec.~\ref{sec:pri}).
\item An alternative formulation of our equations has been
proposed. It has been shown that our equations coincide with the free 
energy variation principle ( Sec.~\ref{sec:alternative}). 
\item  The solution for the two--point correlation (Greens) function
depending on temperature $T$ has been analyzed qualitatively
not cutting the perturbation series.
It includes the low--temperature
solution at $r_0 \to -\infty$ in the case of scalar order--parameter
field~(Sec.~\ref{sec:low1}), as well as the general solution of
$n$--component vector model at $T \to T_c$~(Sec.~\ref{sec:Asbel}).
\item Based on our diagrammatic equations, the
asymptotic long--wave ($k \to 0$) behavior of the transverse
and longitudinal Greens functions below $T_c$ has been analyzed in
Secs.~\ref{sec:leading} to~\ref{sec:join}. This analysis shows that
$G_{\perp}({\bf k}) \simeq a \, k^{-\lambda_{\perp}}$ and
$G_{\parallel}({\bf k}) \simeq b \, k^{-\lambda_{\parallel}}$
with exponents $d/2<\lambda_{\perp}<2$ and $\lambda_{\parallel}
=2 \lambda_{\perp} -d$, and with universal ratio $b M^2/a^2$
is the physical solution of our equations at the spatial
dimensionality $2<d<4$, which coincides with the asymptotic
solution at $T \to T_c$ as well as with
the known rigorous results discussed in Sec.~\ref{sec:cran}
and the non--pertubative
renormalization group arguments provided in Sec.~\ref{sec:rg}.
It is confirmed also by the Monte Carlo simulations in
Sec.~\ref{sec:test}.
Formally, the results of the perturbative RG theory
are recovered at $\lambda_{\perp}=2$. However,
we have disproven the conventional statement of this theory
that $\lambda_{\perp}=2$ is the exact result (Sec.~\ref{sec:cran}).
\item
Monte Carlo simulations in 3D $XY$ model has been performed
(Sec.~\ref{sec:test}) to test the exponent $\rho$, describing
the behavior of
the magnetization $M(h)-M(+0) \sim h^{\rho}$ in small external
fields $h \to 0$ below $T_c$. The simulation results
confirm the universality of this exponent as well as
our prediction $1/2 < \rho <1$, and are incompatible with
the value $\rho=1/2$ of the Gaussian spin wave theory
supported by the perturbative RG analysis.

\end{enumerate}

\section*{Appendix}

Here we study the thermodynamic limit of 
$\Sigma({\bf k},\zeta)$, starting with ${\bf k}={\bf 0}$. 
For simplicity, we consider only the case of scalar 
order--parameter--field, i.~e., $n=1$. The extension to the $n$--component 
case is trivial: the fixed zero--vectors always refer to the longitudinal 
component. According to the definition of $\Sigma({\bf 0},\zeta)$
(cf.~Eqs.~(\ref{eq:ee}) to~(\ref{eq:two})), 
this quantity obeys a selfconsistent diagrammatic equation
$$
\Sigma({\bf 0},\zeta) \simeq \dgn + \dgbn
\eqno{(A1)}
$$
where the block \four represents a resummed perturbation series of 
all connected diagrams of this kind, which do not contain parts like
\selfe and/or \wblock and cannot be split in two as follows \mbox{\fourt.}
At $\zeta=1$, all connected diagrams with four fixed outer lines 
and vectors ${\bf k}_1$, ${\bf k}_2$, ${\bf k}_3$,  
${\bf k}_4= -{\bf k}_1-{\bf k}_2-{\bf k}_3$ represent
the perturbation sum of the cumulant four--point correlation function
$$
G_c \left({\bf k}_1,{\bf k}_2,{\bf k}_3 \right)=
G \left({\bf k}_1,{\bf k}_2,{\bf k}_3 \right) \eqno{(A2)}
$$
$$
-G \left({\bf k}_1,{\bf k}_2 \right) G \left({\bf k}_3,{\bf k}_4 \right)
-G \left({\bf k}_1,{\bf k}_3 \right) G \left({\bf k}_2,{\bf k}_4 \right)
-G \left({\bf k}_1,{\bf k}_4 \right) G \left({\bf k}_2,{\bf k}_3 \right)
$$
where $G \left({\bf k}_1,{\bf k}_2 \right) \equiv 
\delta_{{\bf k}_1,-{\bf k}_2} G({\bf k}_1)$ is the two--point
correlation function. In principle, the diagrams of
$G_c \left({\bf k}_1,{\bf k}_2,{\bf k}_3 \right)$ can be grouped
like those of $\Sigma({\bf k},\zeta)$ following~\cite{K1}. It implies 
the summation over the chains of blocks \mbox{\selfe,} yielding skeleton
diagrams with respect to the solid lines, followed by the summation 
over the chains of blocks \mbox{\ssb,} yielding skeleton diagrams
with respect to the waved lines. We remind that the skeleton diagrams are
those connected diagrams where the true correlation function 
$G({\bf k})$ is related to the solid lines and the dashed lines inside 
the diagrams are replaced by the waved lines. Besides, the skeleton
diagrams do not contain parts \selfe and \mbox{\wblock.} 
In such a way, the perturbation sum of 
$G_c \left({\bf 0},{\bf 0},{\bf 0} \right)$ is almost the same as
\fourzz  with the only difference that it includes also the
diagrams like {\mbox \fourtz.}
It leads to the equation for $G_c \left({\bf 0},{\bf 0},{\bf 0} \right)$,
$$
G_c \left({\bf 0},{\bf 0},{\bf 0} \right)
= \fourzz + \fourtz \;, \eqno{(A3)}
$$
in which the diagram blocks are calculated at $\zeta=1$.
 In analogy to $G({\bf 0}) \simeq M^2 V$ (cf.~eq.~(\ref{eq:Gbel})), 
the zero-vector term $V^{-2} G \left({\bf 0},{\bf 0},{\bf 0} \right)$ 
represents the constant (non--decaying) contribution $M^4$ to the real--space 
four--point correlation function below $T_c$. According 
to~(A2), it means that
$$
G_c \left({\bf 0},{\bf 0},{\bf 0} \right) = -2 M^4V^2  
\hspace{5ex} \mbox{at} \hspace{3ex} V \to \infty \;. \eqno{(A4)}
$$
At the first step we take into account only those contributions
where ${\bf k}={\bf 0}$ vectors are related to the connecting solid lines
inside the last diagram in~(A3). Then, together 
with~(A4), (A1), and~(\ref{eq:wl}),
we have a system of selfconsistent equations yielding
$$
\Sigma({\bf 0},1) \propto V  \eqno{(A5)}
$$
with two possible values of the proportionality factor
$- \left(5 \pm \sqrt{17} \right) \, uM^4$.
One of possibilities is that the additional terms with non--zero 
wave vectors related to the connecting solid lines in~(A3) 
give a negligible corretion at $V \to \infty$, as consistent
with the following idea.
 The four--point function $G \left( {\bf k},-{\bf k},{\bf 0} \right)$
behaves like $M^2V \, G({\bf k})$ below $T_c$, as it follows
from a simple consideration of the limit where
two points of the real--space four--point correlation function
are infinitely distant. However, this term cancellates in the cumulant average
$G_c \left( {\bf k},-{\bf k},{\bf 0} \right)$ at ${\bf k \ne \bf 0}$ 
entering the equation,  analogous to~(A3), 
for the block \mbox{\fourkz.} 
As a result, a selfconsistent solution exists where this
block is vanishingly small as compared to $V$ and Eq.~(A5) remains correct.
Nevertheless, other kind of selfconsistent solutions cannot be excluded
where all terms on the right--hand side of Eq.~(A3) and of 
similar equations for the blocks \fourkz  and \fourkq  
are compatible. The latter is possible if the block in~(A3), 
having four fixed zero--vectors,
diverges as $V^{2+\mu}$ with $\mu \ge 0$, whereas the above mentioned 
blocks with non--zero vectors ${\bf k}$ and ${\bf q}$ diverge like $V^{1+\mu}$
and $V^{\mu}$, respectively. In this case 
$\Sigma({\bf 0},1)$ is proportional to $V^{1+\mu}$.
 
Contrary to the  case of ${\bf k}={\bf 0}$, the thermodynamic limit exists 
for the function $\Sigma({\bf k},1)$ at ${\bf k} \ne {\bf 0}$, as it
can be found easily by a suitable grouping of the divergent terms 
in~(\ref{eq:ee}). Only those
terms can diverge at $V \to \infty$ which contain insertions with $2m$
outer solid zero-vector lines, like \mbox{\fourz,} \mbox{\sixz,}
etc. In an ordinary case $m$ factors $G({\bf 0}) \simeq M^2V$
related to $m$ solid lines with fixed ${\bf k}={\bf 0}$ vectors are 
compensated by
a removal of $m$ integrations over wave vectors, as consistent with the well
known rule\linebreak $\sum_{\bf k} \to V (2 \pi)^{-d} \int d {\bf k}$. 
This condition
is violated if the constraints ${\bf k}={\bf 0}$ are not independent.
The sum of all wave
vectors coming out from any node is zero. As a consequence, this property
holds also for any block. Therefore only $2m-1$ constraints 
${\bf k}={\bf 0}$ for $2m$ outer solid lines of the above discussed
insertions are independent, i.~e., only $2m-1$ integrations are removed.
As a result, any such insertion provides a diverging factor $\propto V$
for the resulting diagram unless this factor is compensated by  
vanishing waved line with fixed ${\bf k}={\bf 0}$. 
Due to these properties, the constraints ${\bf k}={\bf 0}$ for $2m$
solid lines connecting two parts of a diagram like \blockn  also are not
independent, but this situation is possible only for the diagrams of
$\Sigma({\bf 0},\zeta)$. 

Thus, some diagrams containing specific insertions with
$2m$ outer zero--vector solid lines are divergent. At the same time,
all such insertions with $2m$ outer lines represent the perturbation sum
of the $2m$--point cumulant correlation function with all zero arguments. 
In general, the $2m$--point cumulant 
$G_c^{(2m)} \left({\bf k}_1,{\bf k}_2, \ldots , {\bf k}_{2m-1} \right)$
is defined as
$$
G_c^{(2m)} \left({\bf k}_1,{\bf k}_2, \ldots , {\bf k}_{2m-1} \right)
=G^{(2m)} \left({\bf k}_1,{\bf k}_2, \ldots , {\bf k}_{2m-1} \right)
-S^{(2m)} \left({\bf k}_1,{\bf k}_2, \ldots , {\bf k}_{2m-1} \right) \;,
\eqno{(A6)}
$$
where $G_c^{(2m)}$ is given by resummed connected diagrams, i.~e., a
diagram block with $2m$ fixed outer solid lines, whereas $S^{(2m)}$ represents 
the sum over all possible splitings of this block in smaller parts.
These functions contain only $2m-1$ independent arguments (wave vectors of
outer solid lines), since the sum over all $2m$ wave vectors is zero. 
The correlation function 
$G^{(2m)} \left({\bf 0},{\bf 0}, \ldots , {\bf 0} \right)$
is related to the non--decaying part $M^{2m}$ of 
the real--space $2m$--point correlation function and, thus
(according to the Fourier transformation), is $M^{2m}V^m$ at $V \to \infty$.
From~(A6) we find that 
$G_c^{(2m)} \left({\bf 0},{\bf 0}, \ldots , {\bf 0} \right)$
is proportional to $M^{2m}V^m$.
In such a way, we can resummate the specific zero-vector insertions
to replace the perturbation sums by corresponding $2m$--point cumulants. 
It is then straightforward to
see that any inserted $2m$--point zero-vector cumulant does not cause a
divergence at $V \to \infty$, but only renormalize the original diagram by 
a constant factor. Namely,
$m$ solid zero-vector lines of the original diagram are broken to insert
the cumulant block, thus replacing the previous factor 
$[G({\bf 0})]^m \simeq M^{2m}V^m$ of these zero--vector lines 
with the cumulant value $Q_m \, M^{2m}V^m$, where $Q_m$ is a constant. 

Not only the function $\Sigma({\bf k},\zeta)$, but also the derivative
$\partial \Sigma({\bf q},\zeta) /\partial G({\bf k})$ is relevant to
our analysis. This derivative is represented by diagrams obtained by
breaking one solid line with wave vector ${\bf k}$ in the diagrams of 
$\Sigma({\bf q},\zeta)$, removing the corresponding factor $G({\bf k})$.
Let us consider those diagrams of 
$\partial \Sigma({\bf q},1) /\partial G({\bf k})$ with ${\bf k} \ne {\bf 0}$,
containing the zero--vector--cumulant insertions,
where the broken solid line does not belong to any of the inserted
blocks. The diagrams including resummed blocks of this kind and the 
corresponding ``normal'' diagrams whith no insertions differ merely by 
constant factors. Namely, as in the case of $\Sigma({\bf k},1)$, any insertion
of a resummed $2m$--point cumulant block with $2m$ outer zero--vector lines 
gives a constant factor $Q_m$ at $V \to \infty$.   

Below we prove that the derivative $\partial Q_m / \partial G({\bf k})$ 
vanishes at $V \to \infty$. As a result,
the corresponding terms where solid line is broken inside a 
zero--vector--cumulant block do not give an extra contribution to 
$\partial \Sigma({\bf q},1) /\partial G({\bf k})$.
Quantity $M^{2m}V^m \, \partial Q_m / \partial G({\bf k})$, is represented
by the skeleton diagrams of 
$G_c^{(2m)} \left({\bf 0},{\bf 0}, \ldots , {\bf 0} \right)$
in which one solid line with vector ${\bf k}$ is broken in two, removing the 
factor $G({\bf k})$. If factors $G({\bf k})$ and
$G(-{\bf k}) \equiv G({\bf k})$ are restored for both parts of the broken 
line, we obtain the diagrams of 
$G_c^{(2m+2)} \left({\bf k},-{\bf k},{\bf 0}, \ldots , {\bf 0} \right)$.
To simplify the further notation we shall replace the above set of arguments
with one argument ${\bf k}$. Thus, a resummation of these diagrams yields
$$
M^{2m}V^m \,  [G({\bf k})]^2 \, \partial Q_m / \partial G({\bf k}) = 
G_c^{(2m+2)}({\bf k}) \;. \eqno{(A7)}
$$
According to the physical arguments we have
$$
G^{(2m)}({\bf 0})=M^{2m}V^m \;, \eqno{(A8)}
$$
$$
G^{(2m+2)}({\bf k}) = G({\bf k}) \, M^{2m}V^m  \eqno{(A9)}
$$
at $V \to \infty$.
Eq.~(A8) represents the condition for the non--decaying part $M^{2m}V^m$
of the $2m$--point real--space correlation function, whereas~(A9)
describes the limit where $2m$ points of the $(2m+2)$--point function
are infinitely distant.
By setting $m=2$ in~(A6) and taking into account~(A9), we obtain at 
${\bf k} \ne {\bf 0}$
$$ 
M^{-2}V^{-1}G_c^{(4)}({\bf k}) = G({\bf k})-G({\bf k}) \, 
G({\bf 0})M^{-2}V^{-1}=0 
\hspace{5ex} \mbox{at} \hspace{3ex} V \to \infty \;. \eqno{(A10)}
$$
We can prove by induction over $\ell$ that 
$M^{-2\ell}V^{-\ell}G_c^{(2\ell+2)}({\bf k})=0$
holds at $V \to \infty$ for any $\ell \ge 1$ and ${\bf k} \ne {\bf 0}$. 
Eq.~(A10) means that it holds at $\ell=1$. If it holds for
$\ell < m$, then the only relevant terms in the equation~(A6) for
the $(2m+2)$--point cumulant are
$$
G_c^{(2m+2)}({\bf k}) = G^{(2m+2)}({\bf k})-
G({\bf k}) \, G_c^{(2m)}({\bf 0}) - G({\bf k}) \, S^{(2m)}({\bf 0}) \;,
\eqno{(A11)}
$$
which yield vanishing result for $M^{-2m}V^{-m}G_c^{(2m+2)}({\bf k})$
according to Eq.~(A6) for zero--vector cumulants, as well as (A8) and (A9).
According to~(A7), the latter means that quantity
$\partial Q_m / \partial G({\bf k})$ vanishes (at ${\bf k} \ne {\bf 0}$)
in the thermodynamic limit $V \to \infty$. 

However, if it would not vanish, then it would produce
a contribution to the derivative
$\partial \Sigma({\bf q},1) /\partial G({\bf k})$
which is by a factor $V$ larger than the ordinary terms $\sim V^{-1}$. 
It means that not only the leading behavior, but also the corrections to 
finite--size scaling in~(A8) and~(A9) could play some role.
At $n=1$ these corrections are exponentially small, as consistent with the
known exponential decay of the real--space correlation functions in Ising 
model below $T_c$. The latter is true also at $n>1$ for the longitudinal
component of the correlation functions at finite values
of the amplitude $\delta$ of the symmetry--breaking term in~(\ref{eq:H1}),
since in this case Hamiltonian~(\ref{eq:H1}) 
looses the rotational symmetry and, therefore, 
belongs to the Ising universality class.
 Since $\delta$ is a continuous parameter, the exponential decay of 
correlations will be observed at large enough distances $x$ for any 
arbitrarily small, but finite value of $\delta$. 
It means that the corrections are exponentially small in our limit
$\lim\limits_{\delta \to 0} \lim\limits_{L \to \infty}$, where $L$ is
the linear size of the system. There is no any contradiction
with the power--like decay of correlations we found at $n>1$, since
this decay refers to the limit where $x \to \infty$ and, simultaneously,
$x/L \to 0$ hold (i.~e., we find the thermodynamic limit at any given
nonzero wave vector).

Thus, we finally arrive to the conclusion that corrections to~(A8) 
and~(A9) are exponentially small in $L=V^{1/d}$ and, therefore, 
$\partial Q_m / \partial G({\bf k})=0$ holds at 
${\bf k} \ne {\bf 0}$ with a high enough accuracy, i.~e.,  
quantities $Q_m$ can be treated as pure constants providing no 
extra contributions due to their variations.
As discussed before, it means that the inserted zero--vector--cumulants
merely renormalize by constant factors the ``normal'' terms in the 
diagram expansion of $\Sigma({\bf k},1)$ and
$\partial \Sigma({\bf q},1) /\partial G({\bf k})$.
 Since our analysis is not based on specific values of
expansion coefficients, this renormalization cannot
affect our qualitative conclusions. It means that the general
scaling form of $\Sigma({\bf k},1)$ and   
$\partial \Sigma({\bf q},1) /\partial G({\bf k})$
is correctly predicted by the simplified
analysis which ignores the zero--vector--cumulant insertions.

Our foregoing consideration of $\Sigma({\bf k},\zeta)$ is restricted to
$\zeta=1$ due to a technical reason that the relation to cumulant
correlation functions with a certain physical meaning is known
only for this case. 
Since $\zeta$ is a continuous parameter, we believe that the above discussed 
properties remain true for all values of interest, i.~e., 
$\zeta \in [0;1]$. 

We can use the alternative method to find quantity $R_i({\bf k})$, as
proposed in Sec.~\ref{sec:alternative}. It is straightforward to check
in each specific case we considered in our paper that Eq.~(\ref{eq:altern}) 
provides the same scaling properties of $R_i({\bf k})$ at 
${\bf k} \ne {\bf 0}$ as the equations~(\ref{eq:Dast}) and~(\ref{eq:R}). 
Regarding the ``normal'' terms, it is easy to verify
(like in Secs.~\ref{sec:leading} -- \ref{sec:correc} and in
Sec.~\ref{sec:low1}) that in the most nontrivial case of 
$n>1$ at $k \to 0$ below $T_c$, as well as at $r_0 \to -\infty$ in
the case of $n=1$, the main contribution comes from all diagrams of
Eq.~(\ref{eq:altern}) in which the waved lines have nonzero wave 
vectors and each second solid line with $i=1$ 
has fixed wave vector ${\bf k}={\bf 0}$ (i.~e., there are no
integrations in the case of $n=1$), 
counting the pair of outer lines as one line.  
This peculiarity does not refer to the
case $T \to T_c$, where all the partial contributions with
$M^0$, $M^2$, $M^4$, etc., are compatible.
The two different modifications of our method obviously give
consistent corrections to scaling: they have the same origin.
The alternative approach [Eq.~(\ref{eq:altern})] does not suffer from
the problems with parameter $\zeta<1$, since we have $\zeta \equiv 1$.
As before, the zero--vector--cumulant insertions merely renormalize
by constant factors the ``normal'' terms.
It proves the statement that the simplified analysis, ignoring 
these insertions, provides correct general
scaling form of the solution for $R_i({\bf k})$ and, therefore,
$G_i({\bf k})$.

\section*{Acknowledgements}

The Monte Carlo simulations have been made and the results
have been discussed
during my stay at
the Physics Department of Rostock University, Germany.

\end{document}